\documentclass[12pt]{article}
\pdfoutput=1
\usepackage[colorlinks,linkcolor=Blue,citecolor=Blue,bookmarks,bookmarksnumbered]{hyperref}
\usepackage[scaled=0.85]{helvet}
\usepackage{amsmath,amssymb,accents,mathrsfs,XoohmE}
\usepackage{graphicx,color}
\usepackage{booktabs}
\usepackage{multirow}
\usepackage{placeins}
\usepackage{amsmath}

\usepackage{XoohmE}

\definecolor{Green}  {rgb}{0.10,0.70,0.10} 
\definecolor{Orange} {rgb}{1.00,0.50,0.15} 
\definecolor{Red}    {rgb}{0.90,0.00,0.12} 
\definecolor{Purple} {rgb}{0.50,0.25,0.55} 
\definecolor{Turque} {rgb}{0.00,0.65,0.85} 
\definecolor{Blue}   {rgb}{0.00,0.00,1.00} 
\definecolor{Magenta}{rgb}{1.00,0.00,1.00} 
\definecolor{Gold}   {rgb}{1.00,0.75,0.25} 
\definecolor{Seaweed}{rgb}{0.01,0.24,0.09} 
\definecolor{Brown}  {rgb}{0.43,0.26,0.32} 
\definecolor{grey1}  {rgb}{0.20,0.20,0.20} 
\definecolor{grey2}  {rgb}{0.40,0.40,0.40} 
\definecolor{grey3}  {rgb}{0.60,0.60,0.60} 
\definecolor{grey4}  {rgb}{0.80,0.80,0.80} 
\definecolor{grey5}  {rgb}{0.90,0.90,0.90} 
\def\C#1#2{{\ifcase#1\or
             \color{Green}\or \color{Orange}\or \color{Red}\or
              \color{Purple}\or \color{Turque}\or \color{Blue}\or
               \color{Magenta}\or \color{Gold}\or \color{Seaweed}\or
                \color{Brown}\or\color{grey1}\or\color{grey2}\or
                 \color{grey3}\else\color{grey4}\fi#2}}

\definecolor{Slate} {rgb}{0.00,0.45,0.55}

\def\rI{{\rm I}}
\def\rJ{{\rm J}}
\def\rK{{\rm K}}
\def\rL{{\rm L}}

\def\hi{{\hat\imath}}
\def\hj{{\hat\jmath}}
\def\hk{{\hat{k}}}

\def\fracm#1#2{\hbox{\large{${\frac{{#1}}{{#2}}}$}}}

\def\vCent#1{\vcenter{\hbox{\hss#1\hss}}}
\def\be{\begin{equation}}
\def\ee{\end{equation}}
\newcommand{\bea}{\begin{eqnarray}}
\newcommand{\eea}{\end{eqnarray}}
\newcommand{\ena}{\end{eqnarray}}

\def\pp{{\mathchoice
          {
              \kern 1pt%
              \raise 1pt
              \vbox{\hrule width5pt height0.4pt depth0pt
                    \kern -2pt
                    \hbox{\kern 2.3pt
                          \vrule width0.4pt height6pt depth0pt
                          }
                    \kern -2pt
                    \hrule width5pt height0.4pt depth0pt}%
                    \kern 1pt
           }
            {
              \kern 1pt%
              \raise 1pt
              \vbox{\hrule width4.3pt height0.4pt depth0pt
                    \kern -1.8pt
                    \hbox{\kern 1.95pt
                          \vrule width0.4pt height5.4pt depth0pt
                          }
                    \kern -1.8pt
                    \hrule width4.3pt height0.4pt depth0pt}%
                    \kern 1pt
            }
            {
              \kern 0.5pt%
              \raise 1pt
              \vbox{\hrule width4.0pt height0.3pt depth0pt
                    \kern -1.9pt  
                    \hbox{\kern 1.85pt
                          \vrule width0.3pt height5.7pt depth0pt
                          }
                    \kern -1.9pt
                    \hrule width4.0pt height0.3pt depth0pt}%
                    \kern 0.5pt
            }
            {
              \kern 0.5pt%
              \raise 1pt
              \vbox{\hrule width3.6pt height0.3pt depth0pt
                    \kern -1.5pt
                    \hbox{\kern 1.65pt
                          \vrule width0.3pt height4.5pt depth0pt
                          }
                    \kern -1.5pt
                    \hrule width3.6pt height0.3pt depth0pt}%
                    \kern 0.5pt
            }
        }}

\def\mm{{\mathchoice
                       {
                             \kern 1pt
               \raise 1pt    \vbox{\hrule width5pt height0.4pt depth0pt
                                  \kern 2pt
                                  \hrule width5pt height0.4pt depth0pt}
                             \kern 1pt}
                       {
                            \kern 1pt
               \raise 1pt \vbox{\hrule width4.3pt height0.4pt depth0pt
                                  \kern 1.8pt
                                  \hrule width4.3pt height0.4pt depth0pt}
                             \kern 1pt}
                       {
                            \kern 0.5pt
               \raise 1pt
                            \vbox{\hrule width4.0pt height0.3pt depth0pt
                                  \kern 1.9pt
                                  \hrule width4.0pt height0.3pt depth0pt}
                            \kern 1pt}
                       {
                           \kern 0.5pt
             \raise 1pt  \vbox{\hrule width3.6pt height0.3pt depth0pt
                                  \kern 1.5pt
                                  \hrule width3.6pt height0.3pt depth0pt}
                           \kern 0.5pt}
                       }}

\def\ad{{\kern0.5pt
                   \alpha \kern-5.05pt \raise5.8pt\hbox{$\textstyle.$}\kern
0.5pt}}

\def\bd{{\kern0.5pt
                   \beta \kern-5.05pt \raise5.8pt\hbox{$\textstyle.$}\kern
0.5pt}}

\def\qd{{\kern0.5pt
                   q \kern-5.05pt \raise5.8pt\hbox{$\textstyle.$}\kern
0.5pt}}
\def\Dot#1{{\kern0.5pt
     {#1} \kern-5.05pt \raise5.8pt\hbox{$\textstyle.$}\kern
0.5pt}}

\catcode`@=11
\def\un#1{\relax\ifmmode\@@underline#1\else
        $\@@underline{\hbox{#1}}$\relax\fi}
\catcode`@=12

\def\a{\alpha}
\def\b{\beta}
\def\c{\chi}
\def\d{\delta}
\def\e{\epsilon}

\def\i{\iota}
\def\j{\psi}
\def\k{\kappa}
\def\l{\lambda}

\def\o{\omega}

\def\r{\rho}
\def\s{\sigma}
\def\t{\tau}

\def\L{\Lambda}
\def\O{\Omega}
\def\P{\Pi}

\def\S{\Sigma}

\def\ca{{\cal A}}

\def\cf{{\cal F}}

\def\dslash{\not{\hbox{\kern-2pt $\partial$}}}
\def\Dslash{\not{\hbox{\kern-4pt $D$}}}
\def\pslash{\not{\hbox{\kern-2.3pt $p$}}}
 \newtoks\slashfraction
 \slashfraction={.13}
 \def\slash#1{\setbox0\hbox{$ #1 $}
 \setbox0\hbox to \the\slashfraction\wd0{\hss \box0}/\box0 }
 
\def\kcr{{\hbox{\ro \char'170}}}                
\def\ktl{{\hbox{\ro \char'170}}}        
\def\ktr{{\hbox{\ro \char'170}}}        
\def\kbl{{\hbox{\ro \char'170}}}        
\def\kbr{{\hbox{\ro \char'170}}}        

\def\plpl{\raise-2pt\hbox{$\raise3pt\hbox{$_+$}\hskip-6.67pt\raise0.0pt
\hbox{$^+$}\hskip 0.01pt$}}
\def\mimi{\raise-2pt\hbox{$\raise3pt\hbox{$_-$}\hskip-6.67pt\raise0.0pt
\hbox{$^-$}\hskip 0.01pt$}} 

\def\bo{{\raise.15ex\hbox{\large$\Box$}}}               
\def\TH{{\raise.2ex\hbox{$\displaystyle \bigodot$}\mskip-4.7mu \llap H \;}}
\def\face{{\raise.2ex\hbox{$\displaystyle \bigodot$}\mskip-2.2mu \llap {$\ddot
        \smile$}}}                                      

\def\dt#1{\on{\hbox{\bf .}}{#1}}                
\def\Dot#1{\dt{#1}}

\def\Tilde#1{\widetilde{#1}}                    
\def\Hat#1{\widehat{#1}}                        
\def\leftrightarrowfill{$\mathsurround=0pt \mathord\leftarrow \mkern-6mu
        \cleaders\hbox{$\mkern-2mu \mathord- \mkern-2mu$}\hfill
        \mkern-6mu \mathord\rightarrow$}
\def\dvec#1{\vbox{\ialign{##\crcr
        \leftrightarrowfill\crcr\noalign{\kern-1pt\nointerlineskip}
        $\hfil\displaystyle{#1}\hfil$\crcr}}}           
\def\dt#1{{\buildrel {\hbox{\LARGE .}} \over {#1}}}     

\def\fracm#1#2{\hbox{\large{${\frac{{#1}}{{#2}}}$}}}
\def\sfrac#1#2{{\vphantom1\smash{\lower.5ex\hbox{\small$#1$}}\over
        \vphantom1\smash{\raise.4ex\hbox{\small$#2$}}}} 
\def\bfrac#1#2{{\vphantom1\smash{\lower.5ex\hbox{$#1$}}\over
        \vphantom1\smash{\raise.3ex\hbox{$#2$}}}}       
\def\afrac#1#2{{\vphantom1\smash{\lower.5ex\hbox{$#1$}}\over#2}}    

\let\bm\relax
\newcommand{\bm}[1]{{\boldsymbol{#1}}}

\def\ad{{\dot{\alpha}}}
\def\bd{{\dot{\beta}}}

 \font\rOpe=cmsy10                        
 \def\ktl{{\hbox{\rOpe\char'170}}}        
 \def\kbl{{\hbox{\rOpe\char'170}}}        
 \def\kcr{{\reflectbox{\rOpe\char'170}}}        
 \def\ktr{{\reflectbox{\rOpe\char'170}}}        
 \def\kbr{{\reflectbox{\rOpe\char'170}}}        
 \def\Border{\vbox{\hsize0pt
        \setlength{\unitlength}{1mm}
        \newcount\xco
        \newcount\yco
        \xco=-21
        \yco=12
        \begin{picture}(0,0)(-7.5,0)
        \put(\xco,\yco){$\ktl$}
        \advance\yco by-1
        {\loop
        \put(\xco,\yco){$\kcr$}
        \advance\yco by-2
        \ifnum\yco>-240
        \repeat
        \put(\xco,\yco){$\kbl$}}
        \xco=170
        \yco=12
        \put(\xco,\yco){$\ktr$}
        \advance\yco by-1
        {\loop
        \put(\xco,\yco){$\kcr$}
        \advance\yco by-2
        \ifnum\yco>-240
        \repeat
        \put(\xco,\yco){$\kbr$}}
        \put(-19.5,13){\scalebox{.6065}{%
         University of Maryland Center for String and Particle  Theory \&\ Physics Department%
        |University of Maryland Center for String and Particle  Theory \&\ Physics Department}}
        \put(-19.5,-241.5){\scalebox{.5835}{%
         ****University of Maryland * Center for String and
         Particle  Theory* Physics Department****University of Maryland *Center
        for String and Particle  Theory* Physics Department}}
        \end{picture}
        \par\vskip-8mm}}
\definecolor{UMred}{rgb}{.9,.05,.2}
\definecolor{HUblue}{rgb}{.0,.3,.7}

\definecolor{Red}    {rgb}{0.90,0.00,0.12} 
\definecolor{Blue}   {rgb}{0.00,0.00,1.00} 
\definecolor{Green}  {rgb}{0.10,0.70,0.10} 
\definecolor{Turque} {rgb}{0.00,0.65,0.85} 
\definecolor{Orange} {rgb}{1.00,0.50,0.15} 
\definecolor{Magenta}{rgb}{1.00,0.00,1.00} 
\definecolor{Gold}   {rgb}{1.00,0.75,0.25} 
\definecolor{Seaweed}{rgb}{0.01,0.24,0.09} 
\definecolor{Purple} {rgb}{0.50,0.25,0.55} 
\definecolor{Brown}  {rgb}{0.43,0.26,0.32} 
\definecolor{grey1}  {rgb}{0.20,0.20,0.20} 
\definecolor{grey2}  {rgb}{0.40,0.40,0.40} 
\definecolor{grey3}  {rgb}{0.60,0.60,0.60} 
\definecolor{grey4}  {rgb}{0.80,0.80,0.80} 
\definecolor{grey5}  {rgb}{0.90,0.90,0.90} 
\def\C#1#2{{\ifcase#1\or
             \color{Red}\or \color{Green}\or \color{Blue}\or\
              \color{Turque}\or \color{Orange}\or \color{Magenta}\or 
               \color{Gold}\or \color{Seaweed}\or \color{Purple}\or
                \color{Brown}\or\color{grey1}\or\color{grey2}\or
                 \color{grey3}\else\color{grey4}\fi#2}}

\definecolor{Slate} {rgb}{0.00,0.45,0.55}

\newdimen\parshift\parshift=\parindent
\catcode`@=11
 \long\def\@footnotetext#1{\insert\footins{\reset@font\footnotesize
           \interlinepenalty\interfootnotelinepenalty\splittopskip%
            \footnotesep\splitmaxdepth\dp\strutbox\floatingpenalty\@MM%
             \hsize\columnwidth\addtolength{\hsize}{-2\parindent}
              \@parboxrestore\protected@edef\@currentlabel%
              {\csname p@footnote\endcsname\@thefnmark}%
                \color@begingroup%
                 \@makefntext{\rule\z@\footnotesep\ignorespaces#1%
                  \@finalstrut\strutbox}%
                \color@endgroup}}
 \long\def\@makefntext#1{\hglue\parshift%
           \vbox{\noindent\baselineskip=11pt plus.5pt minus.5pt\hb@xt@0em{\hss\@makefnmark\kern1pt}#1}}
\catcode`@=12

\newskip\humongous \humongous=0pt plus 1000pt minus 1000pt
\def\caja{\mathsurround=0pt}
\def\eqalign#1{\,\vcenter{\openup2\jot \caja
        \ialign{\strut \hfil$\displaystyle{##}$&$
        \displaystyle{{}##}$\hfil\crcr#1\crcr}}\,}
\newif\ifdtup
\def\panorama{\global\dtuptrue \openup2\jot \caja
        \everycr{\noalign{\ifdtup \global\dtupfalse
        \vskip-\lineskiplimit \vskip\normallineskiplimit
        \else \penalty\interdisplaylinepenalty \fi}}}

\makeatletter
\def\section{\@startsection{section}{1}{\z@}
        {3ex plus-1ex minus-.2ex}{1pt plus1pt}{\large\sf\bfseries\boldmath}}
\def\subsection{\@startsection{subsection}{2}{\z@}
         {1.5ex plus-1ex minus-.2ex}{0.01pt plus1pt}{\sf\slshape}}
\def\subsubsection{\@startsection{subsubsection}{3}{\z@}
          {1.5ex plus-1ex minus-.2ex}{0.01pt plus0.2pt}{\sf\boldmath}}
\def\paragraph{\@startsection{paragraph}{4}{\z@}
           {.75ex \@plus.5ex \@minus.2ex}{-2mm}{\sf\bfseries\boldmath}}
\makeatother

 \allowdisplaybreaks
 \seceq

\begin{document}

\thispagestyle{empty}
%
\noindent{\small
\hfill{HET-1761  \\ 
$~~~~~~~~~~~~~~~~~~~~~~~~~~~~~~~~~~~~~~~~~~~~~~~~~~~~~~~~~~~~~~~~~$
$~~~~~~~~~~~~~~~~~~~~~~~~~~~~~~~~~~~~~~~~~~~~~~~~~~~~~~~~~~~~~~~~~$
{}
}
\vspace*{8mm}
\begin{center}
{\large \bf
Adinkras From Ordered Quartets of   \\[2pt]
 $\bm {\rm BC{}_4}$
Coxeter Group Elements and Regarding \\[6pt] 
Another Gadget's 1,358,954,496 Matrix 
Elements}   \\   [12mm]
{\large {
S.\ James Gates, Jr.,\footnote{gatess@wam.umd.edu}$^{a}$}
Lucas Kang,\footnote{lucas$_-$kang@brown.edu}$^{a}$
David S. Kessler,\footnote{3.14159.david@gmail.com}$^{b}$ 
and Vadim Korotkikh\footnote{va.korotki@gmail.com}$^{c}$
}
\\*[12mm]
\emph{
\centering
$^{a}$Department of Physics, Brown University,
\\[1pt]
Box 1843, 182 Hope Street, Barus \& Holley 545,
Providence, RI 02912, USA 
\\[12pt]
$^{b}$
Amherst  Center  for  Fundamental  Interactions,  Department  of  Physics,  
\\[1pt]
University  of Massachusetts, Amherst, MA 01003, USA
\\[12pt]
and
\\[12pt] 
$^c$Department of Physics,
University of Maryland, \\[-2pt]
4150 Campus Dr., College Park, MD 20472,  USA
}
 \\*[28mm]
{ ABSTRACT}\\[4mm]
\parbox{142mm}{\parindent=2pc\indent\baselineskip=14pt plus1pt
A Gadget, more precisely a scalar Gadget, is defined as a mathematical calculation
acting over a domain of one or more adinkra graphs and whose range is a real
number.  A 2010 work on the subject of automorphisms of adinkra graphs, implied
the existence of multiple numbers of Gadgets depending on the number of colors
under consideration.  For four colors, this number is two.  In this work, we verify the
existence of a second such Gadget and calculate (both analytically and via explicit
computer-enabled algorithms) its 1,358,954,496 matrix elements over 36,864
minimal valise adinkras related to the Coxeter Group BC${}_4$.}
 \end{center}
\vfill
\vfill
\clearpage

\section{Introduction}

Our study of ``Garden Algebras'' \cite{GRana1,GRana2} and  ``adinkras'' \cite{adnk1} 
has opened multiple pathways for understanding the relation of supersymmetry, in 
most unexpected ways, to topics in mathematics.   Such a demonstration
appeared in the recent examples \cite{adnkGEO1,adnkGEO2} connecting adinkras 
to algebraic geometry as the latest  following a host of others \cite{4Grp,CTa,YZ,adnkM1,adnkM2,codes1,codes2,codes3}.

As adinkras are bipartite graphs, this opens access to methods (e.g.\ \cite{Grf0}) 
associated with graph theory \cite{Grf1,Grf2,Grf3,Grf4,Grf5} as an additional tools to 
study supersymmetry theory.  Indeed, there is a enormously large and rapidly expanding 
literature that emphasizes the importances of networks and graphs across fields 
as diverse as biology, chemistry, computer science, engineering, medicine, neuroscience, and 
physics as noted in the work of \cite{GCG}.  This has prompted at least one call 
to identify all this activity as heralding the emergence of a new field called ``network 
science'' \cite{NS}.  From this vantage, the study of adinkras constitutes a 
network scientific approach to the study of the representation theory of spacetime 
supersymmetry.

In a previous work \cite{permutadnk} motivated by Garden Algebras  and  adinkras, an 
information technology (IT) computer-enabled algorithm was applied to the Coxeter 
Group BC${}_4$ \cite{CXgrp} which contains 384 = $4! \, \times \, 2^4$ elements that 
each describes a signed permutation.  The result of this application was it provided a 
calculational proof of a previously unknown mathematical theorem.  The Coxeter Group 
BC${}_4$ can be regarded as the union of 96 disjoint ``tetrads," subsets ${\widehat 
q}{}_{[1]}$ $\cdots$ ${\widehat q}{}_{[96]}$, i.\ e.
\be
{\rm {BC}}{}_4 ~=~ {\widehat q}{}_{[1]} \,  \cup \,  {\widehat q}{}_{[2]} \,  
\cup \cdots \, \cup \,  {\widehat q}{}_{[96]}  ~=~ \prod_{i = 1}^{96} \, \cup \,
 {\widehat q}{}_{[i]}  ~~~,
\ee
where each tetrad defines a 1D, $N$ = 4 supersymmetry representation as the tetrad 
elements $ \left\{  {\bm {\rm L}}_1, {\bm {\rm L}}_2, {\bm {\rm L}}_3, {\bm {\rm L}}_4 \right\}$
can be chosen to satisfy the Garden Algebra.  An obvious question raised by this result
is whether and how does it extend to other Coxeter Groups?

Given a specified tetrad with elements $ \left\{  {\bm {\rm L}}_1, {\bm {\rm L}}_2, {\bm {
\rm L}}_3, {\bm {\rm L}}_4 \right\}$, by considering all possible replacements $ {\bm {\rm 
L}}_\rI\,$ $\to$ $\pm$ $ {\bm {\rm L}}_\rI\,$ for any fixed value of the subscript I, one
obtains $2^4$ = 16 other tetrads.  We call this type of replacement a ``flip'' of the tetrad.  
Also given the specific tetrad one can re-order the elements within the tetrad.  As each 
tetrad contains four elements, there are 4! = 24 such re-orderings.  We refer to such 
re-orderings as ``flops.''  Taking the flips and flops together, we realize that they constitute 
another realization of the the Coxeter Group BC${}_4$!   When the ordering of the elements 
in tetrads is important that also increases the number of possibilities.  In the work of \cite
{adnkKorrals}, it was shown how to use a bi-quaternionic basis together with the flips,
and the flops, to efficiently describe this large system of possibly independent representations.

By these arguments, one can conclude there are a maximum of 4! $\times$ $2^4$ 
$\times$ 96 = 36,864 1D, $N$ = 4 valise supersymmetry representations related to 
the Coxeter Group BC${}_4$.  It is therefore useful to introduce an index that indicates 
which of these 36,864 representations we wish to specify.  This is done by using a notation 
${\bm {\rm L}}{}_{\rI}^{(\cal R)} $ (when necessary) where the index ${(\cal R)}$ (called 
the ``representation label") takes on values of 1, $\dots$,  36,864. This number can be 
reduced.  For example, if ${\widehat q}{}_{[i]}$ = - ${\widehat q}{}_{[j]}$ with $i$ $\ne$ 
$j$, the two tetrads can be said to describe the same representation for many purposes.

While the valise adinkra representation index ($\cal R$) takes on values from one to 36,864
for the minimal BC${}_4$ valise adinkras, until recently the range of this index for
adinkras with more colors and/or more nodes was not known.  A recent work by Zhang 
\cite{YXZ} has presented a theorem that determines the equivalent number to 36,864 
for any valise adinkra with d nodes (open and closed) and $N$ colors where both
d and $N$ are arbitrary positive integers.

As was noted in the work of \cite{adnk4DCLS}, 4D, Lorentz symmetry and $\cal N$ = 1
supersymmetry, determine the number of minimal off-shell representations to be ten.  By 
use of 4D Hodge duality, this number can be reduced to five.  These five representations 
``live'' in a four dimensional ``weight space-like'' setting characterized by four integers ${
\rm p}_{ ({\widehat {\cal R}})}$, ${\rm q}_{({\widehat {\cal R}})}$, ${\rm r}_{({\widehat {\cal 
R}})}$, and ${\rm s}_{({\widehat {\cal R}})}$.   If one assigns a Euclidean metric for these
coordinates, the supermultiplets in the weight-like space all reside on the surface of a four 
dimensional unit sphere.

One of the 4D, $\cal N$ = 1 supermultiplets lives on the ${\rm s}_{({\widehat {\cal R}})}
$-axis. We can think of this supermultiplet along the ${\rm s}_{({\widehat {\cal R}})} \, \ne$ 
0 as sitting at the apex of a pyramid which has a square as its base.  The remaining four 
supermultiplets live in a three-space orthogonal to the ${\rm s}_{({\widehat {\cal R
}})}$-axis with one of each of the supermultiplets at half of the vertices of a tetrahedron.   

Of the four supermultiplets located at half of the vertices of the tetrahedron, two are 
related to one another via parity exchanges of their coordinates according to ${\rm 
p}_{ ({\widehat {\cal R}})}$ $\to$ $- \, {\rm p}_{ ({\widehat {\cal R}})}$ and ${\rm q}_{({\widehat 
{\cal R}})}$ $\to$ $- \,{\rm q}_{({\widehat {\cal R}})}$.  So in a sense there are only three 
minimal 4D, $\cal  N$ = 1 supermultiplets which we may call the ``chiral supermultiplet'' 
(the one with ${\rm  s}_{({\widehat {\cal R}})}$ $\ne$ 0), the ``vector supermuliplet'' and 
the ``tensor supermultiplet'' (these two along with their parity doubles reside at the 
vertices of a tetrahedron).

Theories that possess 4D, $\cal N$ = 1 spacetime supersymmetry can, via various
reduction techniques, yield 1D, $N$ = 4 theories.  We have used this fact to create 
a method \cite{ENUF2,G-1} for extracting the Garden Algebras and associated 
adinkras for any four dimensional supermultiplet.  Use of such reduction procedures
lead from the three 4D, $\cal N$ = 1 supermultiplets to three adinkras along with their
tetrads...among a sea of 36,864 possible adinkras.  So the question is which adinkras
are selected?  The answer to this depends on the conventions used to write the
4D, $\cal N$ = 1 spacetime supermultiplet as well as the details of the conventions
used in the reduction.

It has been long our suggestion that one should identify an equivalence class
among the adinkras so that the choice of conventions used to write the 4D, $\cal 
N$ = 1 spacetime supermultiplet as well as the details of the conventions
used in the reduction are irrelevant to all members in the equivalence class.
Thus, one is faced with the issue of defining these classes.  This is where the
notion of 4D, $\cal N$ = 1 ``Gadgets''  enter.

Owing to the fact that supercharges transform bosons into fermions and vice-versa,
the notion of a eigenfunction or eigenvector for the supercharges cannot be
defined.  As the usual path to roots and weights of representations precedes along
with the concepts of eigenvectors and eigenvalues, this route cannot be traversed
here.  In the works of \cite{adnk4DCLS,adnk4dgdgt}, it was shown that operators
quadratic in the supercharges when evaluated on the fermionic fields of the
supermultiplets allow for the emergence of numbers that can play the role of the 
eigenvalues.  These are precisely the four integers ${\rm p}_{ ({\widehat {\cal R}})}$, 
${\rm q}_{({\widehat {\cal R}})}$, ${\rm r}_{({\widehat {\cal R}})}$, and  ${\rm s}_{({
\widehat {\cal R}})}$ mentioned above.  The task of evaluating the 4D, $\cal N$ = 1 
``Gadget'' over the (at most ten or minimum of three) supermultiplet representations 
is not a terribly daunting task.
 
On the other hand, as the reduction process injects the images of these 4D, $\cal N$ = 1 
supermultiplets into the sea of 36,864 adinkras, the equivalent task of evaluating 
the analog of the Gadget defined on adinkras amounts to evaluating all 1,358,954,496 
matrix elements in a 36,864 $\times$ 36,864 matrix over the space indexed by ($\cal 
R$).  As shown with the work of \cite{adnkBiLL}, modern computer-enabled algorithms 
permit the explicit evaluation of all matrix elements.  The results are encouraging as 
about 82\% of the matrix elements vanish.  

Even while the adinkra Gadget of \cite{adnkBiLL} was being proposed, prior work
\cite{adnkUWA} indicated for all adinkras with an even number of colors, there are
more similar quantities that can be defined.  Thus, the prediction of a ``second Gadget'' 
is the subject of this current work.  We confirm its existence, and determine all of its
1,358,954,496 matrix elements by two separate methods.
 
The first approach is an analytical one.  It is based on private communications with 
mathematicians (Charles Doran, Jordan Kostiuk \cite{CDJK}, and Yan X. Zhang
\cite{YZ2}) who have been investigating the possibility to generalize the Gadget
in the work \cite{adnk4dgdgt}.  Their approach suggested a simplification in
evaluating Gadgets based on use of the ``$\ell$ and $\Tilde \ell$'' parameters
in one of the two methods proposed to calculate the Gadget in \cite{adnk4dgdgt}.

The second chapter reviews the definition of the first Gadget ${{\cal G}} [  ({ {\cal R}}) , 
( {\cal R}^{\prime}) ]{}_{(1)}$, taking particular note of its two distinct forms expressed 
as (a.) traces over products of matrices and, (b.) as a quadratic form over the ``$\ell$ 
and $\Tilde \ell$'' parameters defined previously.  It is shown how the $\ZZ_2$ valued 
function $\chi_{\rm o}({\cal R})$ \cite{G-1} may be used to express the Gadget in the 
form of a product involving a smaller quadratic form.

In the third chapter, using these analytical expressions relating the two Gadgets we 
are able to predict the range and frequency of the values of the matrix elements of the 
second Gadget ${{\cal G}} [  ({ {\cal R}}) , ( {\cal R}^{\prime}) ]{}_{(2)}$ based on $\chi{
}_{\rm o}({\cal R})$ and the first Gadget ${{\cal G}} [  ({ {\cal R}}) , ( {\cal R}^{\prime}) ]
{}_{(1)}$ over both the small BC${}_4$ library (introduced in \cite{adnk4dgdgt}) as 
well as over all 36,864 values of $({\cal R})$.

The fourth chapter presents the results of the calculations for the 1,358,954,496 matrix 
elements when the second Gadget is evaluated over the small BC${}_4$ library (where 
it defines as 96 $\times$ 96 matrix) and over the entire range of the representation label 
$({\cal R})$ where it is a 36,864 $\times$ 36,864 matrix.  The results are written in the 
form of frequency tables of the entries of the in the second Gadget as well as by use of 
the ``Summary of the Gadget'' holomorphic functions (as introduced in \cite{adnkBiLL}).  It 
turns out that the second Gadget possesses the exact same set of entries whose values 
equal zero as the first Gadget, possesses some entries that take on the new value of -1
(compared to the first Gadget), and re-arranges the frequencies of the entries of the first 
Gadget.

The fifth chapter presents the easily access graphical presentation of the
results for both the first and second Gadgets considered over the small
BC${}_4$ library.  This chapter contains scalable pdf images for each
Gadget that allows the explicit values of the matrix elements to be
determined via a color code scheme.

We next turn in the sixth chapter to the description of three set of codes, the 
first two written in Python and the final in MATLAB.  The MATLAB code 
was created as a ``fast check'' for the arguments made for analytical derived 
results over the small BC${}_4$ library.  After agreement was shown here, 
the Python code was run over both the BC${}_4$ small library as well as
over the complete 36,864  values of the entire $({\cal R})$ library.  Agreement 
 was shown for both the small library results as well as the analytically 
 derived results.  Finally, the two different Python codes were created 
 independently to act as additional quality checks on the generation of
 the 2,717,908,992 matrix elements generated for ${{\cal G}} [  ({ {\cal R}}) , 
 ( {\cal R}^{\prime}) ]{}_{(1)}$ and ${{\cal G}} [  ({ {\cal R}}) , ( {\cal R}^{\prime}) ]
{}_{(2)}$ over the complete range of the 36,864 adinkras.

Our Python codes will be provided free to any interested reader as we 
support open access research in our domain.  This can be provided by making 
a simple request via e-mail to any of the authors.  Alternately, any interested party 
upon clicking the links at:

(a.) https://github.com/vkorotkikh/SUSY-Adinkra-Gadget2, and

(b.) https://github.com/lk11235/SUSY$_-$AdinkraGadget2$_-$parallel
\vskip.02in \noindent
on-line can initiate a download of these codes.
 
 The seventh chapter contains our conclusions.

\newpage

\section{Optimizing The First Gadget}
\label{s2a}

By use of a set of Feynman-like rules, any adinkra representation $(\cal R)$ 
leads to a set of $ {\bm {\rm L}}^{(\cal R)}_\rI $ and ${\bm {\rm R}}^{(\cal R)}_\rI$ matrices 
that satisfy the  ``Garden Algebra'' conditions,
\be { \eqalign{
 (\,{\rm L}^{(\cal R)}_\rI\,)_i{}^\hj\>(\,{\rm R}^{(\cal R)}_\rJ\,)_\hj{}^k ~+~ (\,{\rm L
 }^{(\cal R)}_\rJ\,)_i{}^\hj\>(\,{\rm R}^{(\cal R)}_\rI\,)_\hj{}^k &= 2\,\d_{\rI\rJ}\,\d_i{}^k
 ~~,\cr
(\,{\rm R}^{(\cal R)}_\rJ\,)_\hi{}^j\>(\, {\rm L}^{(\cal R)}_\rI\,)_j{}^\hk ~+~ (\,{\rm 
 R}^{(\cal R)}_\rI\,)_\hi{}^j\>(\,{\rm L}^{(\cal R)}_\rJ\,)_j{}^\hk  &= 2\,\d_{\rI\rJ}\, \d_\hi{
 }^\hk~~,  \cr
~~~(\,{\rm R}^{(\cal R)}_\rI\,)_\hj{}^k\,\d_{ik} = (\,{\rm L}^{(\cal R)}_\rI\,)_i{
}^\hk\,\d_{\hj\hk}&~~,
}}\label{GarDNAlg2}
\ee
and for a specified adinkra representation $(\cal R)$, these can be used to define 
two additional sets of matrices.   We have given the name of ``bosonic holoraumy 
matrices'' and  ``fermionic holoraumy matrices,'' respectively, to these other sets denoted 
by $\bm {V_{\rI\rJ}}^{(\cal R)}$ and $\bm{{\Tilde V}}_{\rI\rJ}{}^{(\cal R)}$ 
\cite{adnk4dgdgt}, and defined via the equations
\be { \eqalign{
 (\,{\rm L}^{(\cal R)}_\rI\,)_i{}^\hj\>(\,{\rm R}^{(\cal R)}_\rJ\,)_\hj{}^k ~-~ (\,{\rm L
 }^{(\cal R)}_\rJ\,)_i{}^\hj\>(\,{\rm R}^{(\cal R)}_\rI\,)_\hj{}^k &= i\, 2\,  (V_{\rI\rJ}^{(\cal 
 R)})_i{}^k~~,\cr
 (\,{\rm R}^{(\cal R)}_\rI\,)_\hi{}^j\>(\, {\rm L}^{(\cal R)}_\rJ\,)_j{}^\hk ~-~ (\,{\rm 
 R}^{(\cal R)}_\rJ\,)_\hi{}^j\>(\,{\rm L}^{(\cal R)}_\rI\,)_j{}^\hk  &= i \, 2\,  
 ({\Tilde V}_{\rI\rJ}^{(\cal R)})_\hi{}^\hk ~~.
}}\label{GarDVs}
\ee
Alternately, if we suppress all the matrix indices above, these equations take the
forms
\be { \eqalign{
{\bm {\rm L}}^{(\cal R)}_\rI \,  {\bm {\rm R}}^{(\cal R)}_\rJ ~+~ 
{\bm {\rm L}}^{(\cal R)}_\rJ \,  {\bm {\rm R}}^{(\cal R)}_\rI 
&=  2\,\d_{\rI\rJ}\,   \bm{{\rm I}}
 ~~,\cr
{\bm {\rm R}}^{(\cal R)}_\rI \,  {\bm {\rm L}}^{(\cal R)}_\rJ ~+~ 
{\bm {\rm R}}^{(\cal R)}_\rJ \,  {\bm {\rm L}}^{(\cal R)}_\rI 
& =   2\,\d_{\rI\rJ}\,   \bm{{\rm I} }  ~~,\cr
 {\bm {\rm L}}^{(\cal R)}_\rI  ~=~ [ {\bm {\rm R}}^{(\cal R)}_\rI  ]{}^t   &~~.
}}\label{GarDVsNid1}  
\ee
(where the superscript $t$ indicates matrix transposition while
$\bm {\rm I}$ denotes the identity matrix) for the first set and 
\be { \eqalign{
{\bm {\rm L}}^{(\cal R)}_\rI \,  {\bm {\rm R}}^{(\cal R)}_\rJ ~-~ 
{\bm {\rm L}}^{(\cal R)}_\rJ \,  {\bm {\rm R}}^{(\cal R)}_\rI 
&= i\, 2\,   \bm{{V}}_{\rI\rJ}{}^{(\cal R)}
 ~~,\cr
{\bm {\rm R}}^{(\cal R)}_\rI \,  {\bm {\rm L}}^{(\cal R)}_\rJ ~-~ 
{\bm {\rm R}}^{(\cal R)}_\rJ \,  {\bm {\rm L}}^{(\cal R)}_\rI 
&= i \, 2\,  \bm{{\Tilde V}}_{\rI\rJ}{}^{(\cal R)}
 ~~,
}}\label{GarDVsNid2}  
\ee
for the second set.

Given two adinkras representations denoted by $({ {\cal R}})$  and $( {\cal R}^{
\prime})$ (which possess $N$ colors, and $d$ open nodes \& $d$ closed nodes) 
along with their associated fermionic holoraumy matrices $\bm{{\Tilde V}}_{\rI\rJ}
{}^{(\cal R)}$ and $\bm{{\Tilde V}}_{\rI\rJ}{}^{({\cal R}^{\prime})}$ we form a 
scalar, ``the gadget value'' between two representations $({ {\cal R}})$  and $(
{\cal R}^{\prime})$ defined by
\be
{{\cal G}} [  ({ {\cal R}}) , ( {\cal R}^{\prime}) ] ~=~ \left[ \, {1 \over {~ N \, (N-1) \, 
d_{\min}(N)}}  \,   \right] \, \sum_{\rI , \, \rJ} \,{\rm {Tr}} \,  \left[ \, \bm{{\Tilde V}}_{\rI\rJ}
{}^{(\cal R)}  \, \bm{{\Tilde V}}_{\rI\rJ}{}^{({\cal R}^{\prime})}  \right]  ~~~,
\label{Gdgt1}
\ee
where we exclude the cases of $N$ = 0, 1 in (\ref{Gdgt1}).  The function
$d_{\min}(N)$ is given by
\be
d_{\min}(N)=\begin{cases}
{~~}2^{\frac{N-1}{2}}\,~,&N\equiv 1,\,7 $~~~~~~~\,~~~$ \bmod{(8)}\\
{~~}2^{\frac{N}{2}}{~~~~},&N\equiv 2,\,4,\,6 $~~~~~~~$ \bmod{(8)}\\
{~~}2^{\frac{N+1}{2}}~\, ,&N\equiv 3,\,5 $~~~~\,~~~~~~$ \bmod{(8)}\\
{~~}2^{\frac{N - 2}{2}}~\, ,&N\equiv  8 $~~~~~~~~~~~~~~$ \bmod{(8)}\\
\end{cases}
{~~~~~~,~~~~~~~~~}
\label{eqn:dmin}
\ee
and where only the case of $N$ = 0 (i.e.\ no supersymmetry) in (\ref{eqn:dmin}) 
is excluded in this definition. It can be seen from (\ref{eqn:dmin}) that for all values 
of $N$ $>$ 1, the number of nodes {\em {must}} be a power of two.  For every adinkra 
\cite{adnk1,Grf1} based on the Coxeter Group ${\rm {BC}}_4$, the ${\rm 
L}^{(\cal R)}_\rI\,$ and ${\rm R}^{(\cal R)}_\rI$ matrices \cite{GRana1,GRana2} 
must have four colors  (${\rm I} = 1,\dots, 4$), four open nodes ($i = 1,\dots, 4$), and 
four closed nodes ($\hk = 1,\dots, 4$). 

As a consequence of (\ref{GarDNAlg2}), the  $\bm{{\Tilde V}}_{\rI\rJ}{}^{(\cal R)}$ and 
$\bm{{\Tilde V}}_{\rI\rJ}{}^{({\cal R}^{
\prime})}$ matrices are actually elements of the so(4) algebra.  Due to this, they may 
be expanded in the ``$\a$-$\b$'' basis.  In other words, the six matrices defined by 
 \be {
\begin{array}{cccc}
&{\bm {\a}}{}^{\,\Hat 1} ~=~ {\bm \s}^2 \otimes {\bm \s}^1 ~~, & ~~ {\bm {\a}}{}^{\,\Hat 2} ~=~  {\bm 
{\rm I}}{}_{2 \times 2}  \otimes {\bm \s}^2  ~~, &  
~~{\bm {\a}}{}^{\,\Hat 3}  ~=~ {\bm \s}^2 \otimes {\bm \s}^3   ~~, \\
&{\bm {\b}}{}^{\,\Hat 1}  ~=~  {\bm \s}^1 \otimes {\bm \s}^2~~, & ~~ {\bm {\b}}{}^{\,\Hat 2} ~=~ {\bm 
\s}^2 \otimes  {\bm {\rm I}}{}_{2 \times 2} ~~, &  ~~{\bm {\b}}{}^{\,\Hat 3} ~=~  {\bm \s}^3 
\otimes {\bm \s}^2  ~~, \\ \end{array}
} \label{aLbE}
 \ee
 can be chosen as the basis over which to expand $ ({\Tilde V}_{\rI\rJ}^{(\cal 
 R)})_\hi{}^\hk$ in the form
\be   \eqalign{ {~~~~~~~}
 {\bm {\Tilde V}{}_{\rI\rJ}^{(\cal R)}} ~=~ &\Big[ \,\ell^{({\cal R}){\Hat 1}}_{\rI\rJ}\, 
 {\bm {\a}}{}^{\,\Hat 1}  \, + \,  \ell^{({\cal R}){\Hat 2}}_{\rI\rJ}\, {\bm {\a}}{}^{\,\Hat 2} 
 \,+\,  \ell^{({\cal R}){\Hat 3} }_{\rI\rJ}\, {\bm {\a}}{}^{\,\Hat 3}    \, \Big]     
 ~+~  \Big[ \,    {{\Tilde \ell}^{(\cal R)}}_{\rI\rJ}{
 }^{\Hat 1}\,  {\bm {\b}}{}^{\,\Hat 1}  \,+\, \, {{\Tilde \ell}^{(\cal R)}}_{\rI\rJ}{}^{\Hat 2}\, 
  {\bm {\b}}{}^{\,\Hat 2} 
 \,+\, {{\Tilde \ell}^{(\cal R)}}_{\rI\rJ}{}^{\Hat 3}\,  {\bm {\b}}{}^{\,\Hat 3}    \, \Big]   ~~~,
}  \label{Veq}
\ee
written in terms of a set of 36 coefficients $\ell^{({\cal R}){\Hat 1}}_{\rI\rJ}$, $\ell^{({\cal R}){\Hat 
2}}_{\rI\rJ}$, $\ell^{({\cal R}){\Hat 3}}_{\rI\rJ}$, ${\Tilde \ell}^{({\cal R}){\Hat 1}}_{\rI\rJ}$, ${\Tilde 
\ell}^{({\cal R}){\Hat 2}}_{\rI\rJ}$, and ${\Tilde \ell}^{({\cal R}){\Hat 3}}_{\rI\rJ}$.

The values of the ``Gadget,'' expressed in terms of the ${\ell}$ and ${ {\Tilde \ell}}$ coefficients, 
is defined by a quadratic form
\be
\eqalign{
{ {\cal G}} \left[  ({\cal R}) \, , \,  ({\cal R}^{\prime}) \right]{}_{\ell} ~\equiv ~~~\frc{1}{6}
\sum_{{\Hat a} } \, {\Big [} ~ 
&{\ell}_{1 \, 2}^{\,({\cal R}) \Hat{a}} \,  {\ell}_{1 \, 2}^{\, ({\cal R}^{\prime}) \Hat{a}}  ~+~ 
{\ell}_{1 \, 3}^{\,({\cal R}) \Hat{a}} \,  {\ell}_{ 1 \, 3}^{\, ({\cal R}^{\prime}) \Hat{a}} ~+~
{\ell}_{1\, 4}^{\,({\cal R}) \Hat{a}} \,  {\ell}_{1\, 4}^{\, ({\cal R}^{\prime}) \Hat{a}}    ~+~ \cr
&{\ell}_{2 \, 3}^{\, ({\cal R}) \Hat{a}}  {\ell}_{2 \, 3}^{\, ({\cal R}^{\prime}) \Hat{a}} \, ~+~ 
{\ell}_{2 \, 4}^{\,({\cal R}) \Hat{a}} \,  {\ell}_{ 2 \, 4}^{\, ({\cal R}^{\prime}) \Hat{a}} ~+~
{\ell}_{3 \, 4}^{\,({\cal R}) \Hat{a}} \,  {\ell}_{3\, 4}^{\, ({\cal R}^{\prime}) \Hat{a}}   ~+~ \cr
&{\Tilde {\ell}}_{1 \, 2}^{\,({\cal R}) \Hat{a}} \,  {\Tilde {\ell}}_{1 \, 2}^{\, ({\cal R}^{\prime}) \Hat{a}}  ~+~ 
{\Tilde {\ell}}_{1 \, 3}^{\,({\cal R}) \Hat{a}} \,  {\Tilde {\ell}}_{ 1 \, 3}^{\, ({\cal R}^{\prime}) \Hat{a}} ~+~
{\Tilde {\ell}}_{1\, 4}^{\,({\cal R}) \Hat{a}} \,  {\Tilde {\ell}}_{1\, 4}^{\, ({\cal R}^{\prime}) \Hat{a}}    ~+~ \cr
&{\Tilde {\ell}}_{2 \, 3}^{\, ({\cal R}) \Hat{a}}  {\Tilde {\ell}}_{2 \, 3}^{\, ({\cal R}^{\prime}) \Hat{a}}~+~ 
{\Tilde {\ell}}_{2 \, 4}^{\,({\cal R}) \Hat{a}} \,  {\Tilde {\ell}}_{ 2 \, 4}^{\, ({\cal R}^{\prime}) \Hat{a}} ~+~
{\Tilde {\ell}}_{3 \, 4}^{\,({\cal R}) \Hat{a}} \,  {\Tilde {\ell}}_{3\, 4}^{\, ({\cal R}^{\prime}) \Hat{a}}  
~ {\Big ]}
~~~,  } \label{Gdgt2}\ee
and necessarily when $ ({ {\cal R}})$ = $( {\cal R}^{\prime}) $ this is a non-negative
quadratic form.  Due to this property, an angle between $ ({ {\cal R}})$ and $( {\cal R
}^{\prime}) $ can be defined by
\be {
cos \left\{ \theta [({{\cal R}})\,  , \, ({ {\cal R}}^{\prime} )] {}_{\ell} \right\} ~=~
{{{ {\cal G}} [ \, ({ {\cal R}}) , ({ {\cal R}}^{\prime}) \,]{}_{\ell} } \over {~ {\sqrt{{ 
{\cal G}} [ \, ({ {\cal R}}) , ({ {\cal R}}) \, ]{}_{\ell}}} \, {\sqrt{ { {\cal G}} [ \, ({ {\cal 
R}}^{\prime}) , ({ {\cal R}}^{\prime}) ]{}_{\ell}}}~~ } } ~~~,  }   \label{M4gL}
\ee
and clearly when the condition
\be
{ {\cal G}} [ \, ({ {\cal R}}) , ({ {\cal R}}) \,]{}_{\ell} ~=~ 1 ~~~,
\ee
is met, the Gadget value corresponds directly to the cosine of an angle.

In recent works \cite{CDJK,YZ2}, there has been exploited a calculational advantage to 
using a different form of (\ref{Gdgt2}) by making note that for all adinkras constructed 
from elements in ${\rm {BC}}_4$, the
identities
 \be  \eqalign{
\ell^{({\cal R}){\Hat \a}}_{\rI\rJ}  ~=~ \fracm 12 \, {\chi}_{\rm o}({\cal R}) \,  \e^{\rI\rJ \rK \rL}
 \ell^{({\cal R}){\Hat \a}}_{\rK \rL}  ~~~,~~~ {\Tilde \ell}^{({\cal R}){\Hat \a}}_{\rI\rJ}  ~=~ 
 \fracm 12 \, {\chi}_{\rm o}({\cal R}) \,  \e^{\rI\rJ \rK \rL} {\Tilde \ell}^{({\cal R}){\Hat \a}}_{\rK \rL} 
 ~~~,
}  \label{IdX}
\ee
are valid.  In this expression ${\chi}_{\rm o}({\cal R})$ is a function (as first identified
in the work of \cite{G-1}) that {\em {only}} takes on values of either $+1$ or $-1$ for any 
${\rm {BC}}_4$ based adinkras.  Using this, the result in (\ref{Gdgt2}) can be expressed 
as the product of two factors
\be
\eqalign{
{ {\cal G}} \left[  ({\cal R}) \, , \,  ({\cal R}^{\prime}) \right]{}_{\ell} ~\equiv ~\frc{1}{6} \,
 \left[ \,  1\, + \,  {\chi}_{\rm o}({\cal R}) \,  {\chi}_{\rm o}({\cal R}^{\prime}) \, \right] \,
{ {\rm G}}\left[  ({\cal R}) \, , \,  ({\cal R}^{\prime}) \right]{}_{\ell}
~~~.  } \label{Gdgt2A}\ee
where the second factor ${ {\rm G}} \left[  ({\cal R}) \, , \,  ({\cal R}^{\prime}) \right]
{}_{\ell}$ takes the explicit form give by
\be
\eqalign{
{ {\rm G}} \left[  ({\cal R}) \, , \,  ({\cal R}^{\prime}) \right]{}_{\ell} ~= ~&\frc{1}{3} \,  
{\Big \{} \sum_{{\Hat a} } \, {\Big [} ~ {\ell}_{1 \, 2}^{\,({\cal R}) \Hat{a}} \,  {\ell}_{1 \, 
2}^{\, ({\cal R}^{\prime}) \Hat{a}}  ~+~  {\ell}_{1 \, 3}^{\,({\cal R}) \Hat{a}} \,  {\ell}_{
1 \, 3}^{\, ({\cal R}^{\prime}) \Hat{a}} ~+~ {\ell}_{2\, 3}^{\,({\cal R}) \Hat{a}} \,  {\ell
}_{2\, 3}^{\, ({\cal R}^{\prime}) \Hat{a}}  ~+
\cr
&{~~~~}{~~~~~~~~~}{\Tilde {\ell}}_{1 \, 2}^{\,({\cal R}) \Hat{a}} \,  {\Tilde {\ell}}_{1 \, 
2}^{\, ({\cal R}^{\prime}) \Hat{a}}  ~+~ {\Tilde {\ell}}_{1 \, 3}^{\,({\cal R}) \Hat{a}} \,  
{\Tilde {\ell}}_{1 \, 3}^{\, ({\cal R}^{\prime}) \Hat{a}}  ~+~ {\Tilde {\ell}}_{2\, 3}^{\,({
\cal R}) \Hat{a}} \,  {\Tilde {\ell}}_{2\, 3}^{\, ({\cal R}^{\prime}) \Hat{a}}  ~ {\Big ]}    
{\Big\}}   
~~~,  } \label{Gdgt2Bx2}\ee
It should be noted that for all adinkras based on BC${}_4$, either all the $\ell^{({\cal 
R}){\Hat \a}}_{\rI\rJ} $ {\em {or}} the $ {\Tilde \ell}^{({\cal R}){\Hat \a}}_{\rI\rJ}$ coefficient
vanish and it has been noted \cite{CDJK,YZ2} that the ``pre-factor'' $1/2 \left[ \,  1 +   
{\chi}_{\rm o}({\cal R}) \,  {\chi}_{\rm o}({\cal R}^{\prime})   \, \right]$ only takes on values 
of one and zero.  A final comment about (\ref{Gdgt2Bx2}) is the expression for ${ {\rm 
G}} \left[  ({\cal R}) \, , \,  ({\cal R}^{\prime}) \right]{}_{\ell}$ has a form that is identical
to that of a Gadget defined on a space of adinkras with three colors.

Direct calculation further has shown \cite{adnkBiLL} that for all minimal BC${}_4$ based 
valise adinkras, the Gadget only takes on one of four possible values, i.\ e.\ -$1/3$, 0, $
1/3$ or 1.   Accordingly, the only angles that occur between the 36,864 BC${}_4$ based 
adinkras are $arccos(- 1/3) $, $\pi/2$, $arccos(1/3)$, and 0.  This number of angles, i.\ e.
\ four, is about $10^{-16}$ times smaller than if the angles were totally at random distributed 
among the 36,864 BC${}_4$ based adinkras.  Thus, the system of all BC${}_4$ 
based minimal valise adinkras is highly ordered.

Hereafter, we will call this the ``first Gadget" utilizing the convention
\be  \eqalign{
{{\cal G}}{}_{(1)
} [  ({ {\cal R}}) , ( {\cal R}^{\prime}) ]
~&\equiv ~ 
{{\cal G}} [  ({ {\cal R}}) , ( {\cal R}^{\prime}) ] 
 ~~~,
}    \label{Gdgt1a}
\ee
and as this name suggests, there is a second Gadget to be defined shortly.

\newpage
\section{Defining The Second Gadget}
\label{s2bw}

We can define the ``second Gadget" according to the equation 
\be
{{\cal G}}{}_{(2)} [  ({ {\cal R}}) , ( {\cal R}^{\prime}
) ] ~=~  \fracm 1{96} \, \sum_{\rI , \, \rJ ,\, 
\rK , \, \rL} \, \e^{\rI\rJ \rK \rL} \, {\rm {Tr}} \,  \left[ \, 
\bm{{\Tilde V}}_{\rI\rJ}{}^{(\cal R)}  \, \bm{{\Tilde V}}_{\rK\rL}{}^{({\cal R}^{\prime}
)}  \right]  ~~~,
\label{Gdgt2zz}
\ee
based on the use of the Levi-Civita tensor $ \e^{\rI\rJ \rK \rL}$.  The evaluation of this 
over all possible BC${}_4$ related adinkras representations $({ {\cal R}})$, and $({\cal 
R}^{\prime})$ leads to another set of 1,358,954,496 matrix elements.  The existence of 
multiple numbers of Gadget can be seen as an implication of the work of \cite{adnkUWA}.  
There in equation (23) of the work and expression is given for adinkras with arbitrary 
numbers of colors.  The case of four colors is covered in equation (21) of the work.  These
equations also appeared in the work of \cite{G-1} where they were presented in equations 
numbered as (71) and (78) respectively.

We can also note the following identity
\be  \eqalign{
{}
 \sum_{\rI , \, \rJ ,\, \rK , \, \rL} \, 
 \e^{\rI\rJ \rK \rL} \, {\rm {Tr}} \,  \left[ \, \bm{{\Tilde V}}_{\rI\rJ}{}^{(\cal R)}  
 \, \bm{{\Tilde V}}_{\rK \rL }{}^{({\cal R}^{\prime})}  \right]  ~=~ 
 & 4 \,  {\rm {Tr}} \,  \left[ \, \bm{{\Tilde V}_{1 \, 2}}{}^{(\cal R)}  
 \, \bm{{\Tilde V}_{3\, 4}}{}^{({\cal R}^{\prime})}  \right]  \,-\,
 4 \,  {\rm {Tr}} \,  \left[ \, \bm{{\Tilde V}_{1 \, 3}}{}^{(\cal R)}  
 \, \bm{{\Tilde V}_{2\, 4}}{}^{({\cal R}^{\prime})}  \right]    \cr
 &+ \, 4 \,  {\rm {Tr}} \,  \left[ \, \bm{{\Tilde V}_{1 \, 4}}{}^{(\cal R)}  
 \, \bm{{\Tilde V}_{2\, 3}}{}^{({\cal R}^{\prime})}  \right]  \,+\,
 4 \,  {\rm {Tr}} \,  \left[ \, \bm{{\Tilde V}_{2 \, 3}}{}^{(\cal R)}  
 \, \bm{{\Tilde V}_{1 \, 4}}{}^{({\cal R}^{\prime})}  \right]      \cr
 &- 4 \,  {\rm {Tr}} \,  \left[ \, \bm{{\Tilde V}_{2 \, 4}}{}^{(\cal R)}  
 \, \bm{{\Tilde V}_{1\, 3}}{}^{({\cal R}^{\prime})}  \right]  \,+\,
 4 \,  {\rm {Tr}} \,  \left[ \, \bm{{\Tilde V}_{3 \, 4}}{}^{(\cal R)}  
 \, \bm{{\Tilde V}_{1 \, 2}}{}^{({\cal R}^{\prime})}  \right]   ~~,
} \label{2Gdgt2}\ee
which implies further
\be  \eqalign{
{~~~~}
 \sum_{\rI , \, \rJ ,\, \rK , \, \rL} \, 
 \e^{\rI\rJ \rK \rL} \, {\rm {Tr}} \,  \left[ \, \bm{{\Tilde V}}_{\rI\rJ}{}^{(\cal R)}  
 \, \bm{{\Tilde V}}_{\rK \rL }{}^{({\cal R}^{\prime})}  \right]  ~=~ 
 &  16 \sum_{{\Hat a} } \, {\Big [} ~ 
{  {\ell}}_{1 \, 2}^{\,({\cal R}) \Hat{a}} \,  {  {\ell}}_{3 \, 4}^{\, ({\cal R}^{\prime}) 
\Hat{a}} \, \,-\, {  {\ell}}_{1 \, 3}^{\,({\cal R}) \Hat{a}} \,  {  {\ell}}_{ 2 \, 4}^{\, ({\cal 
R}^{\prime}) \Hat{a}}\,+\,{  {\ell}}_{1\, 4}^{\,({\cal R}) \Hat{a}} \,  {  {\ell}}_{2\, 
3}^{\, ({\cal R}^{\prime}) \Hat{a}}  ~~ \,+\,\cr
& {~~~~~~~~~~~~}
{  {\ell}}_{2 \, 3}^{\, ({\cal R}) \Hat{a}}  {  {\ell}}_{1 \, 4}^{\, ({\cal R}^{\prime}) \Hat{a}}~-~ 
{  {\ell}}_{2 \, 4}^{\,({\cal R}) \Hat{a}} \,  {  {\ell}}_{ 1 \, 3}^{\, ({\cal R}^{\prime}) \Hat{a}} \,+\,
{  {\ell}}_{3 \, 4}^{\,({\cal R}) \Hat{a}} \,  {  {\ell}}_{1\, 2}^{\, ({\cal R}^{\prime}) \Hat{a}}  ~~+
  \cr
 &{~~~~~~~~~~~~}
{\Tilde {\ell}}_{1 \, 2}^{\,({\cal R}) \Hat{a}} \,  {\Tilde {\ell}}_{3 \, 4}^{\, ({\cal R}^{\prime
}) \Hat{a}}  ~-~ {\Tilde {\ell}}_{1 \, 3}^{\,({\cal R}) \Hat{a}} \,  {\Tilde {\ell}}_{ 2 \, 4}^{\, 
({\cal R}^{\prime}) \Hat{a}} ~+~ {\Tilde {\ell}}_{1\, 4}^{\,({\cal R}) \Hat{a}} \,  {\Tilde 
{\ell}}_{2\, 3}^{\, ({\cal R}^{\prime}) \Hat{a}}   \,+\, \cr
& {~~~~~~~~~~~~}
{\Tilde {\ell}}_{2 \, 3}^{\, ({\cal R}) \Hat{a}}  {\Tilde {\ell}}_{1 \, 4}^{\, ({\cal R}^{\prime}) 
\Hat{a}} ~-~ {\Tilde {\ell}}_{2 \, 4}^{\,({\cal R}) \Hat{a}} \,  {\Tilde {\ell}}_{ 1 \, 3}^{\, ({\cal 
R}^{\prime}) \Hat{a}} ~+~ {\Tilde {\ell}}_{3 \, 4}^{\,({\cal R}) \Hat{a}} \,  {\Tilde {\ell}}_{1
\, 2}^{\, ({\cal R}^{\prime}) \Hat{a}}  ~ {\Big ]}
~~~,   
} \label{2Gdgt2z}\ee
or more simply the second Gadget takes the alternate form
\be   \eqalign{
{{\cal G}}{}_{(2)} [  ({ {\cal R}}) , ( {\cal R}^{\prime}) ]{}_{\ell} ~=&~ 
\fracm 16 \sum_{{\Hat a} } \, {\Big [} ~ {  {\ell}}_{1 \, 2}^{\,({\cal R}) \Hat{a}} \,  
{  {\ell}}_{3 \, 4}^{\, ({\cal R}^{\prime}) \Hat{a}}  ~-~ {  {\ell}}_{1 \, 3}^{\,({\cal R}) \Hat{a}} \, 
 {  {\ell}}_{ 2 \, 4}^{\, ({\cal R}^{\prime}) \Hat{a}} ~+~ {  {\ell}}_{1\, 4}^{\,({\cal R}) \Hat{a}} 
 \,  {  {\ell}}_{2\, 3}^{\, ({\cal R}^{\prime}) \Hat{a}}    \,+~ \cr
& {~~~~~~~~~~~}
{  {\ell}}_{2 \, 3}^{\, ({\cal R}) \Hat{a}}  {  {\ell}}_{1 \, 4}^{\, ({\cal R}^{\prime}) \Hat{a}} \, ~-~ 
{  {\ell}}_{2 \, 4}^{\,({\cal R}) \Hat{a}} \,  {  {\ell}}_{ 1 \, 3}^{\, ({\cal R}^{\prime}) \Hat{a}} ~+~
{  {\ell}}_{3 \, 4}^{\,({\cal R}) \Hat{a}} \,  {  {\ell}}_{1\, 2}^{\, ({\cal R}^{\prime}) \Hat{a}}  \,~+     \cr
& {~~~~~~~~~~~}
{\Tilde {\ell}}_{1 \, 2}^{\,({\cal R}) \Hat{a}} \,  {\Tilde {\ell}}_{3 \, 4}^{\, ({\cal R}^{\prime}) 
\Hat{a}}  ~-~ {\Tilde {\ell}}_{1 \, 3}^{\,({\cal R}) \Hat{a}} \,  {\Tilde {\ell}}_{ 2 \, 4}^{\, ({\cal 
R}^{\prime}) \Hat{a}} ~+~ {\Tilde {\ell}}_{1\, 4}^{\,({\cal R}) \Hat{a}} \,  {\Tilde {\ell}}_{2
\, 3}^{\, ({\cal R}^{\prime}) \Hat{a}}    ~+~ \cr
& {~~~~~~~~~~~}
{\Tilde {\ell}}_{2 \, 3}^{\, ({\cal R}) \Hat{a}}  {\Tilde {\ell}}_{1 \, 4}^{\, ({\cal R}^{\prime}) 
\Hat{a}}~-~ {\Tilde {\ell}}_{2 \, 4}^{\,({\cal R}) \Hat{a}} \,  {\Tilde {\ell}}_{ 1 \, 3}^{\, ({\cal 
R}^{\prime}) \Hat{a}} ~+~ {\Tilde {\ell}}_{3 \, 4}^{\,({\cal R}) \Hat{a}} \,  {\Tilde {\ell}}_{1
\, 2}^{\, ({\cal R}^{\prime}) \Hat{a}}  ~ {\Big ]}
~~~,
{~}} \label{2Gdgt3}\ee
when expressed in terms of the coefficients $\ell^{({\cal R}){\Hat 1}}_{\rI\rJ}$, $\ell^{(
{\cal R}){\Hat 2}}_{\rI\rJ}$, $\ell^{({\cal R}){\Hat 3}}_{\rI\rJ}$, ${\Tilde \ell}^{({\cal R}){\Hat 
1}}_{\rI\rJ}$, ${\Tilde \ell}^{({\cal R}){\Hat 2}}_{\rI\rJ}$, and ${\Tilde \ell}^{({\cal R}){\Hat 
3}}_{\rI\rJ}$.  Upon comparing (2.7) to (3.4), it can be seen that the only difference
in the two equations is the reversal of the signs for one-third of the terms.

We can also use (\ref{IdX}) to write (\ref{2Gdgt3}) as the product of two factors
\be
\eqalign{
{ {\cal G}}{}_{(2)} \left[  ({\cal R}) \, , \,  ({\cal R}^{\prime}) \right]{}_{\ell} ~= ~&\frc{1}{6} 
\, \left[ \,   {\chi}_{\rm o}({\cal R}) \,  + \,  {\chi}_{\rm o}({\cal R}^{\prime})   \, \right] \, 
{{\rm G}} [   ({ {\cal R}}) , ( {\cal R}^{\prime}) ]{}_{\ell}   
~~~,  } \label{Gdgt2B}\ee
where the second factor is the same as appears in (\ref{Gdgt2A}).  We see that
when $ ({ {\cal R}})$ = $( {\cal R}^{\prime}) $ (see the definition of 
$ \chi{}_{\rm o} $ \cite{G-1}) we have
\be   \eqalign{
{{\cal G}}{}_{(2)} [  ({ {\cal R}}) , ( {\cal R}) ] ~=&~ \chi{}_{\rm o}  ({ {\cal R}})
~~~,
{~}} \label{2Gdgt4}\ee
since the value of the sum taken over any BC${}_4$ related adinkras always yields a 
factor of three.  Thus, the values of $ \chi{}_{\rm o} $ (i.\ e.\ ``chi-oh'') correspond to the 
diagonal entries of the second Gadget.  The ``pre-factor'' $1/2 \left[ \,   {\chi}_{\rm o}({\cal 
R})  +   {\chi}_{\rm o}({\cal R}^{\prime})   \, \right]$ only takes on values of minus one, zero, 
and plus one.  Moreover we note that 
\be
\left[ \,   {\chi}_{\rm o}({\cal R})  \,+\,   {\chi}_{\rm o}({\cal R}^{\prime})   \, \right] ~=~0  ~ 
\leftrightarrow ~ \left[ \, 1 \,+\,  {\chi}_{\rm o}({\cal R}) \, {\chi}_{\rm o}({\cal R}^{\prime})   
\, \right] ~=~ 0 ~~~,
\ee
so that the zeroes of the two pre-factors are exactly the same pairs of representations.
  
Since $ \chi{}_{\rm o} $ is not positive definite,  ${{\cal G}}{}_{(2)} [  ({ {\cal R}}) , ( {\cal 
R}) ]$ cannot be used to define a Euclidean metric on the space of adinkras with the 
property that the angle between two adinkra representations $ ({ {\cal R}})$ and $( 
{\cal R}^{\prime}) $ vanishes when $ ({ {\cal R}})$ = $( {\cal R}^{\prime}) $.  The factors 
of - 1 along the diagonal of ${{\cal G}}{}_{(2)} [  ({ {\cal R}}) , ( {\cal R}^{\prime}) ]$, when
considered as a matrix over the 36,864 BC${}_4$ adinkras, prevents the second Gadget 
from being used in the formula of (\ref{M4gL}) to define a second angle between adinkra 
representations.

From (\ref{Gdgt2A}) it follows that the identity
\be
\fracm 12 \, \left[ \,  1 \,  + \,  {\chi}_{\rm o}({\cal R}) \,  {\chi}_{\rm o}({\cal R}^{\prime})   \, \right] 
 \, {{\cal G}}{}_{(1)} [  ({ {\cal R}}) , ( {\cal R}^{\prime}) ] ~=~  {{\cal G}}{}_{(1)} [  
 ({ {\cal R}}) , ( {\cal R}^{\prime}) ] ~~~,
  \label{Id1}
\ee
should be valid.
Furthermore, 
since 
\be
\left[ \,   {\chi}_{\rm o}({\cal R}) \,  + \,  {\chi}_{\rm o}({\cal R}^{\prime})   \, \right]  \,
\left[ \,   {\chi}_{\rm o}({\cal R}) \,  + \,  {\chi}_{\rm o}({\cal R}^{\prime})   \, \right]  ~=~ 
 2\,  \left[ \, 1 \,  + \,  {\chi}_{\rm o}({\cal R})  \,  {\chi}_{\rm o}({\cal R}^{\prime})   \, 
 \right]  ~~~,
\ee
it follows the identities
\be
 \left[ \,   {\chi}_{\rm o}({\cal R}) \,  + \,  {\chi}_{\rm o}({\cal R}^{\prime})   \, \right] 
 \, {{\cal G}}{}_{(2)} [  ({ {\cal R}}) , ( {\cal R}^{\prime}) ] ~=~ 2  \, {{\cal G}}{}_{(1)} [  
 ({ {\cal R}}) , ( {\cal R}^{\prime}) ]  ~~~,
 \label{Id2}
\ee
\be
{\chi}_{\rm o}({\cal R}) \, {{\cal G}}{}_{(1)} [  ({ {\cal R}}) , ( {\cal R}^{\prime}) ] ~=~   
{\chi}_{\rm o}({\cal R}^{\prime}) \, {{\cal G}}{}_{(1)} [  ({ {\cal R}}) , ( {\cal R}^{\prime}) ] ~=~ 
{{\cal G}}{}_{(2)} [   ({ {\cal R}}) , ( {\cal R}^{\prime}) ]  ~~~,
\label{Id3} 
\ee
should be valid.  The final line above implies a more symmetrical expression
\be
{{\cal G}}{}_{(2)} [   ({ {\cal R}}) , ( {\cal R}^{\prime}) ] 
~=~ \fracm 12 \left[ \, {\chi
}_{\rm o}({\cal R}) \,  + \,  {\chi}_{\rm o}({\cal R}^{\prime}) \, \right] {{\cal G}}{
}_{(1)} [   ({ {\cal R}}) , ( {\cal R}^{\prime}) ] ~~~,
\label{Id4}
\ee
which informs us that knowledge of  the values of ${\chi}_{\rm o}({\cal R})$ and
$ {{\cal G}}{}_{(1)} [   ({ {\cal R}}) , ( {\cal R}^{\prime}) ]$ are sufficient to
determine the values of $ {{\cal G}}{}_{(2)} [   ({ {\cal R}}) , ( {\cal R}^{\prime}) ]$ 
over the entirety of the space of BC${}_4$ related adinkras.

The formula in (\ref{Id4}) predicts the behavior of the second Gadget without 
explicit evaluation of (\ref{Gdgt2zz}).  Since the pre-factor of $ \fracm 12 \left[ 
\, {\chi}_{\rm o}({\cal R})   +   {\chi}_{\rm o}({\cal R}^{\prime}) \, \right] $ only takes on the 
values of 0 and $\pm 1$, this implies that $ {{\cal G}}{}_{(2)} [   ({ {\cal R}}) , ( {\cal 
R}^{\prime}) ]$ can only take on the values of 0, $\pm 1/3$, and $\pm 1$.  Thus, the 
second Gadget possesses a set of entries $-1$ that did not appear in the first 
Gadget. 

All the identities in (\ref{Id1}), (\ref{Id2}), (\ref{Id3}),  and (\ref{Id4}), stem from the
facts that the two Gadgets, when evaluated on BC${}_4$ based adinkras, take the 
forms  
\be  \eqalign{
{{\cal G}}{}_{(1)} [   ({ {\cal R}}) , ( {\cal R}^{\prime}) ]  &=~ \fracm 12 \left[ \,1 \, + \,
{\chi}_{\rm o}({\cal R})  {\chi}_{\rm o}({\cal R}^{\prime}) \, \right] 
{{\rm G}} [   ({ {\cal R}}) , ( {\cal R}^{\prime}) ]{}_{\ell}    ~~~,  \cr
{{\cal G}}{}_{(2)} [   ({ {\cal R}}) , ( {\cal R}^{\prime}) ]  &=~  \fracm 12 \left[ \, {\chi
}_{\rm o}({\cal R}) \,  + \,  {\chi}_{\rm o}({\cal R}^{\prime}) \, \right] 
{{\rm G}} [   ({ {\cal R}}) , ( {\cal R}^{\prime}) ]{}_{\ell}    ~~\,~~,
}  \label{Gdgt2Bx1} \ee
with the factor of ${ {\rm G}} \left[  ({\cal R}) \, , \,  ({\cal R}^{\prime}) \right]{}_{\ell}$
being a common one in both definitions.  We are also to write one more valid
equation of the form
\be
{{\cal G}}{}_{(2)} [   ({ {\cal R}}) , ( {\cal R}^{\prime}) ] 
~=~ \fracm 12 \left\{ \, 
{{\cal G}}{}_{(2)} [   ({ {\cal R}}) , ( {\cal R}) ] 
\,  + \,  
{{\cal G}}{}_{(2)} [   ({ {\cal R}^{\prime}}) , ( {\cal R}^{\prime}) ] 
 \, \right\} {{\cal G}}{
}_{(1)} [   ({ {\cal R}}) , ( {\cal R}^{\prime}) ] ~~~,
\label{Id5}
\ee
for the matrix elements of the two Gadgets.

\newpage
\section{Results For The Second Gadget}
\label{s2bu}

The formula in (\ref{Id4}), without explicit evaluation of (\ref{Gdgt2zz}), also predicts 
the frequencies with which the non-vanishing elements appear.  The frequencies 
of appearances of $- 1/3$, 0, $1/3$, and 1 in the first Gadget  are shown in Table 1
\vskip3pt
\begin{table}[h]
\begin{center}
\footnotesize
\begin{tabular}{|c|c|}
\hline 
~~~ Gadget${}_{(1)}$ Value ~~ & ~~~~Count \\  \hline   \hline  
- 1/3 & ~~\,~\,127,401,984 ~  \\   \hline   
~ 0 & ~\, 1,132,462,080 ~~  \\ \hline   
 ~ 1/3 &  ~~~~\, 84,934,656 ~  \\ \hline   
 ~1 &  ~\,~~\,\, 14,155,776 ~  \\  \hline 
\end{tabular}
\end{center}
\end{table}
$~~~~~~~~~~~~~~~~~~~$ $~$ {\bf {Table}} {\bf {1:}} Frequency of Appearance of  
$ {{\cal G}} {}_{(1)} [   ({ {\cal R}}) , ( {\cal R}^{\prime}) ]$ Elements
\vskip0.1in  \noindent
and the formula of (\ref{Id4}) permits one to predict that the frequency of 
appearances of $\pm 1$, $\pm \fracm 13$, and 0 in the second Gadget 
are given as shown in Table 2. 
\vskip3pt
\begin{table}[h]
\begin{center}
\footnotesize
\begin{tabular}{|c|c|}
\hline 
~~~ Gadget${}_{(2)}$ Value ~~ & ~~~~Count \\  \hline   \hline  
- 1 & ~~\,~\, 7,077,888 ~  \\   \hline  
- 1/3 & ~~\,~\,106,168,320 ~  \\   \hline   
~ 0 & ~\, 1,132,462,080 ~~  \\ \hline   
 ~ 1/3 &  ~~~~\, 106,168,320 ~  \\ \hline   
 ~1 &  ~\,~~\,\, 7,077,888 ~  \\  \hline 
\end{tabular}
\end{center}
\end{table}
$~~~~~~~~~~~~~~~~~~~$ $~$ {\bf {Table}} {\bf {2:}} Frequency 
of Appearance of $ {{\cal G}}
 {}_{(2)} [   ({ {\cal R}}) , ( {\cal R}^{\prime}) ]$ Elements
\vskip0.1in  \noindent

In the work of \cite{adnkBiLL} there was introduced a meromorphic ``Summary of the Gadget'' 
function for the first Gadget which in light of the introduction of the second Gadget we 
now write as
\be \eqalign{
 {\cal S}{}_{{\cal G}(1)} (z) ~&=~ 
{1 \over  {~  z^{p_1}  \, (\, z \,+\, \fracm 13 \,)^{p_2}  \, (\, z \,-\, \fracm 13 \,)^{p_3}
\,   \, (\, z \,-\, 1 \,)^{p_4} ~}}  ~~~,   \cr
& p_1 ~=~ 1,132,462,080 ~~,~~ p_2 ~=~ 127,401,984  ~~,~~  \cr
&p_3 ~=~  84,934,656
 ~~~~\,~~,~~ p_4 ~=~ 14,155,776   ~\,~~.
} \ee
\vskip0.5pt \noindent 
 {{This expression has the following properties:}}
 \vskip0.5pt  $~~~~\,~~~$
(a.) the sum of the exponents equals to the square of
\newline \indent $~~~~~~\,~~~~~~~$ 
36,864, i.e.\ the rank of the first Gadget matrix,
\newline $~~~~~~~~~~~~~$
(b.) the poles of this function are the only non-vanishing
\newline \indent $~~~~~~\,~~~~~~~$  
entries that appear in the first Gadget matrix, and
\newline $~~~~~~~~~~~~~$
(c.) the exponent associated with each pole is the multiplicity  
\newline \indent $~~~~~~~\,~~~~~~$  
with which the value of the pole appears in the
first \newline \indent $~~~~~~~\,~~~~~~$ Gadget matrix.

Due to the results in Table 2, we can write a similar expression for a meromorphic ``Summary 
of the Gadget'' function for the second Gadget 
\be \eqalign{
 {\cal S}{}_{{\cal G}(2)} (z) ~&=~ 
{1 \over  { ~ z^{p_1}  \, (\, z^2 \,-\, \fracm 19 \,)^{q_1}  \, (\, z^2 \,-\, 1 \,)^{q_2} ~
}}  ~~~,   \cr
& q_1 ~=~ 106,168,320  ~~,~~   q_2 ~=~  7,077,888   ~\,~~.
} \ee
\vskip0.5pt \noindent  
This expression has the following properties:
\vskip0.5pt  $~~~~\,~~~$
(a.) the sum $p_1$ + 2($q_1 + q_2$) equals to the square of  
\newline \indent $~~~~~~\,~~~~~~~$ 36,864, i.e.\ the rank of the second 
Gadget matrix, and
\newline $~~~~~~~~~~~~~$
(b.) the poles of this function are the only non-vanishing
\newline \indent $~~~~~~\,~~~~~~~$  entries that appear in the second Gadget matrix.

\newpage
\section{Visual Graph of The Second Adinkra Gadget Values Over The ``Small  
${\rm BC}_4$ Library"}
\label{s2abc}

\indent
$~~~~$ In the work of \cite{adnkBiLL}, we started the demonstration of the structure of
the results found for the first Gadget by giving a visual representation over a tiny
section of the entries of the full 36,864 $\times$ 36,864 symmetrical matrix.  This tiny 
section consisted of an examination of only a 96 $\times$ 96 sub-sector of the
symmetrical matrix. The 96 adinkra representations contained in the ``small ${\rm 
BC}_4$ library" are defined by starting with the Coxeter Group ${\rm BC}_4$, which 
contains 384 elements, followed by forming tetrads.

In equation (5.6) of that work, six collections of permutations are presented.  Each 
of these collections possesses four permutation elements.  Next one takes these
collections and multiplies them by the six collections of Boolean Factors shown in 
Appendix B of \cite{adnkBiLL}.  Since each collection of Boolean Factors contains 
sixteen elements, one ends up with 6 $\times$ 16 = 96 tetrads.  The representation 
of these 96 tetrads described in terms of their $\ell$ and $\Tilde {\ell}$ values is then 
given in Appendix C of \cite{adnkBiLL}.  Finally, one takes these $\ell$ and $\Tilde {\ell}$ 
values and substitutes them in (\ref{2Gdgt3}) of this work to calculate the values
under the second Gadget.  To obtain explicit values, a MATLAB code was created 
(this will be described in more detail later) to evaluate the Second gadget over the 
``small ${\rm BC}_4$ library." 

We can use the concept of the Summary of the Gadget approach but restricted to the 
small ${\rm BC}_4$ library, and denote by ${\cal S}{}_{{\cal G}(1)}^{{\rm BC}_4} (z) $
and ${\cal S}{}_{{\cal G}(2)}^{{\rm BC}_4} (z) $ for the two Gadgets respectively to find
\be \eqalign{ {~~~~~~~~~~}
 {\cal S}{}_{{\cal G}(1)}^{{\rm BC}_4} (z) ~&=~ 
{1 \over  { ~ z^{r_1}  \, (\, z \,+\, \fracm 13 \,)^{2 r_2}  \, (\, z \,-\, 1 \,)^{2 r_3} ~
}}    ~~,~~
 {\cal S}{}_{{\cal G}(2)}^{{\rm BC}_4} (z) ~=~ 
{1 \over  { ~ z^{r_1}  \, (\, z^2 \,-\, \fracm 19 \,)^{r_2}  \, (\, z^2 \,-\, 1 \,)^{r_2} ~
}}  ~~~,   \cr
& r_1 ~=~ 4,608  ~~,~~   r_2 ~=~  640  ~~,~~   r_3 ~=~  1,664  ~\,~~,
} 
\label{s2GdgtBC}
\ee
We note 4,608  + 2(640) +  2(1,664) = 9,216 = (96) $\times$ (96).  Since the order 
of the poles at zero are the same for these two expression, this implies the number 
of zeroes in the two Gadgets is the same.  However, the MATLAB code also verified 
that not only the number of zero entries was the same, it also verified that the locations 
of the 4,608 zeros occurs in the identically corresponding location of each respective 
matrix.

A set of visual predictions that follow from (\ref{Id4}) can be made also.  These
include:

       $~~~~$ (i.) none of the white ``pixels'' are to be modified,

       $~~~~$ (ii.) half of the green ``pixels" will be replaced by
             blue ``pixels," 

       $~~~~$ (iii.) some red ``pixels" are to be replaced by
              black ``pixels,"

       $~~~~$ (iv.) the
              number of red ``pixels" must be the same as 
              the number \newline $~~~~~~~~~~~~~~~~$ of black ``pixels,"

       $~~~~$ (v.) the
              number of green ``pixels'' must be the same as 
              the number 
              \newline $~~~~~~~~~~~~~~~~$ of blue ``pixels," and

       $~~~~$ (vi.) on the diagonal only 48 green ``pixels" and 48
              blue ``pixels" appear.

To efficiently show the results for the 9,216 matrix elements over the ``small ${\rm 
BC}_4$ library," we introduce a color key shown in Table 3 relating colors to 
numerical values.  So instead of showing the numerical values, we can use 
9,216 ``pixels'' in their place.
\newpage
$$
\vCent
{\setlength{\unitlength}{1mm}
\begin{picture}(-20,0)
\put(-48,-40){\includegraphics[width=3.6in]{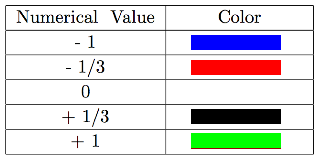}}
\put(-43,-49){{{\bf {Table}} {\bf {3:}}
Color Key For Matrix Elements }}
\end{picture}}
\nonumber
$$
\vskip2.0in  \noindent

This \footnote{The image shown in Fig.\ 1 is taken from the text of \cite{adnkBiLL}.} 
permits the visual representation for all the matrix elements of ${\cal 
G}_{(1)}^{{\rm BC}_4}$ shown in Fig.\ 1,
$$
\vCent
{\setlength{\unitlength}{1mm}
\begin{picture}(-20,0)
\put(-104,-256){\includegraphics[width=8.2in]{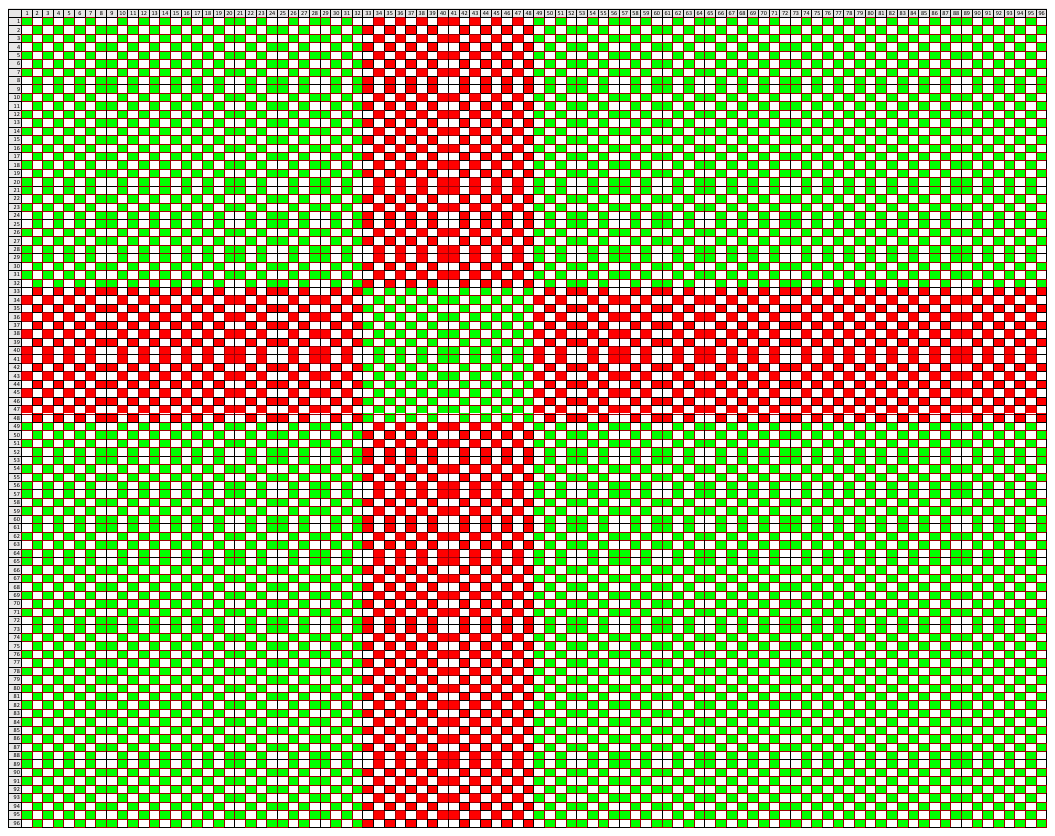}}
\put(-86,-149){{{\bf {Figure}} {\bf {1:}}
Visual Representation of the Values in 
First Gadget Adinkra Representation Matrix }}
\end{picture}}
\nonumber
$$
\newpage \noindent
and as well to illustrate the visual representation of the all matrix elements of 
${\cal G}_{(2)}^{{\rm BC}_4}$ shown in Fig.\ 2.

$$
\vCent
{\setlength{\unitlength}{1mm}
\begin{picture}(-20,0)
\put(-104,-244){\includegraphics[width=8.2in]{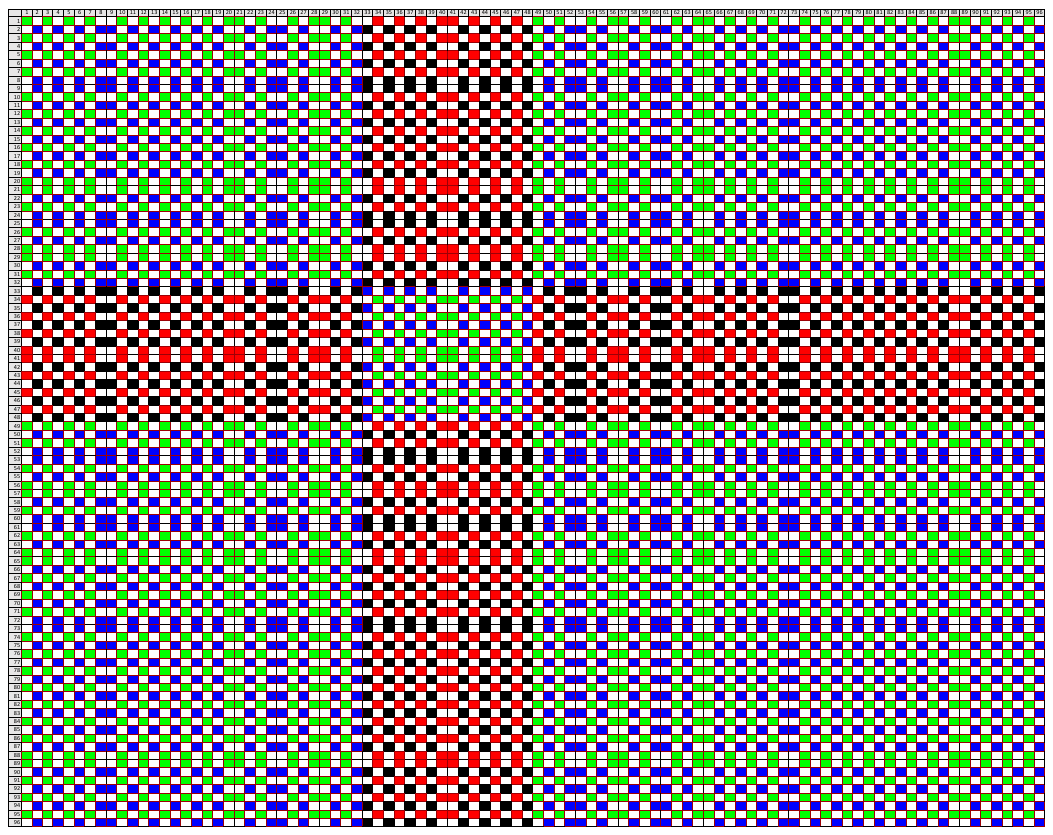}}
\put(-86,-139){{{\bf {Figure}} {\bf {2:}}
Visual Representation of the Values in 
Second Gadget Adinkra Representation Matrix }}
\end{picture}}
\nonumber
$$
$$~~$$
\vskip5.3in
The images in Fig.\ 1 and Fig.\ 2 are scalable pdf files in the native
format of this document which implies it can be magnified as desired
to obtain greater detail.

We can also express the data about the frequency of the various matrix entries
that occur in the ${\rm {BC}}_4$ Small Library Count in tabular form.  This is 
shown in Table 3 and Table 4.

\vskip3pt
\begin{table}[h]
\begin{center}
\footnotesize
\begin{tabular}{|c|c|}
\hline 
~~~ Gadget${}_{(1)}$ Value ~~ & ${\rm {BC}}_4$ Small Library Count \\  \hline   \hline  
 - 1 &  ~\,~~\,\, ~~~~~~~~~~~ 0 ~  \\  \hline
- 1/3 & ~~\,~\, ~~~~~~~~1,280 ~  \\   \hline   
~ 0 & ~\, ~~~~~~~~ ~~\,4,608 ~~  \\ \hline   
 ~ 1/3 &  ~~~~\, ~~~~~~~~~~~~0 ~  \\ \hline   
 ~1 &  ~\,~~\,\, ~~~~~~ 3,328 ~  \\  \hline 
\end{tabular}
\end{center}
\end{table}
$~~~~~~~~~~$ {\bf {Table}} {\bf {3:}} Frequency of Appearance of 
$ {{\cal G}} {}_{(1)} [   ({ {\cal R}}) , ( {\cal R}^{\prime}) ]$ Elements Over Small Library
\newpage

\begin{table}[h]
\begin{center}
\footnotesize
\begin{tabular}{|c|c|}
\hline 
~~~ Gadget${}_{(2)}$ Value ~~ & ${\rm {BC}}_4$ Small Library Count \\  \hline   \hline  
- 1 & ~~\,~\, ~~~ \,~~ 1,664 ~  \\   \hline  
- 1/3 & ~~\,~\, ~~~~~~~~ 640 ~  \\   \hline   
~ 0 & ~\, ~~~~~ ~~~~ 4,608 ~~  \\ \hline   
 ~ 1/3 &  ~~\,~\, ~~~~~~~~ 640 ~   \\ \hline   
 ~1 &  ~~\,~\, ~~~ \,~~ 1,664 ~   \\  \hline 
\end{tabular}
\end{center}
\end{table}
$~~~~~~~~$ $~$ {\bf {Table}} {\bf {4:}} Frequency of Appearance of  
$ {{\cal G}} {}_{(2)} [   ({ {\cal R}}) , ( {\cal R}^{\prime}) ]$ Elements Over Small Library

Having verified (\ref{Id4}) over the 9,216 entries of the small library, we next turn to the
issue of evaluating it over the 1,358,954,496 entries of the entirety of ${\rm {BC}}_4$.
A discussion of the code used for doing this appears in chapter six.

\newpage
\section{The Codes}
\label{codes}

\subsection{Python Code-I Version}

This subsection contains a general description of the Python software code used in
explicitly calculating the matrix elements of ${{\cal G}}{}_{(2)} [  ({ {\cal R}}) , ( {\cal
R}^{\prime}) ] $ over the entirety of 36,864 adinkras as well as for the ${\rm {BC}
}_4$ Small Library.  The code was written in Python 3.5 and executed using Python
version 3.5 but it is also compatible with Python 2.7.

\subsubsection{Creating all ${\rm {BC}}_4$ space Adinkras}
The Python code builds upon earlier developments done in \cite{adnkBiLL}
but with modifications to the code that pertain to the second Gadget calculation
as well as new code for increasing calculation speed and saving the results.
To speed up the Gadget calculation, multiprocessing feature was added and is
utilized within the code. The code also now produces a text output of the results
which can be zip compressed for easier distribution of results.\par
The ${\rm {BC}}_4$ Coxeter group elements are defined by adinkra{\_}nxn{\_}constructor.py
script which creates the 384 L-matrices. These L-matrices serve as the building
blocks of all adinkras, given that any two satisfy conditions of the Garden
algebra equations. Once the 384 L-matrices are created, for each L matrix the
script builds a list of compatible matrices that satisfy Garden Algebra conditions.
Given 384 L-matrices, allowing for all ``color permutations'' in an adinkra there
is a total of 36,864 adinkras with four color, four open node and four close node.

\subsubsection{Calculating Holoraumy Matrices}
The fx{\_}vij{\_}holoraumy.py script is used to calculate the Holoraumy matrices
for each one adinkra of the 36,864. This script can calculate the ${\bm {\tilde {\rm
V}}} $ Fermionic Holoraumy matrices well as the $\ell$ and $\tilde{\ell}$ coefficients
for ${{\bm {\tilde{\rm V}}}}$ matrices. The fermionic{\_}holomats function generates
a set of six ${{\bm {\tilde{\rm V}}}}$ matrices for any given adinkra.

The calc{\_}vij{\_}alphabeta function uses the six ${{\bm {\tilde{\rm V}}}}$ matrices
to calculate the corresponding $\ell$ and $\tilde{\ell}$ coefficient values. Once the
calculation is finished executing the fx{\_}vij{\_}holoraumy.py script returns a list
(a mutable data structure in Python) of calculated $\ell$ and $\tilde{\ell}$ coefficients.
These coefficients and the ${{\bm {\tilde{\rm V}}}}$ matrices will be used in second
Gadget calculations.  The value calculation for the second Gadget is done by the
fx{\_}mpgadgets.py script. This script utilizes the multiprocessing module available
in Python to greatly speed up the Gadget calculation process. The values for the
second Gadget can be calculated via two different ways.

First is taking the sum of all traces of ${{\bm {\tilde{\rm V}}}}$ matrices from
any two adinkras. If $(\mathcal{R})$ and $(\mathcal{R}^\prime)$ are any two adinkras
from the set of 36,864 then the Gadget II app is defined by the equation (\ref{Gdgt2zz}).
This is performed in newgadget{\_}mtraces and newgadget{\_}trace functions.  These
functions are not multithreaded and only return the Gadget II values to caller.
Gadget II can also be calculated by using the $\ell$ or $\tilde{\ell}$ coefficients
of any two $(\mathcal{R})$ and $(\mathcal{R}^\prime)$ adinkras via use of (\ref{2Gdgt3}).

As a check on the consistency of the calculations both methods were used.  No
differences were found between Gadget II values calculated via the Trace method
or the $\tilde{\ell}$ coefficient method.

The advantage to using the $\tilde{\ell}$ coefficients is that it eliminates extensive
matrix dot product calculations and instead simplifies each gadget calculation to
summing up the 1 and -1. This type of Gadget II calculation method is done in the
mp{\_}gadgetcalc{\_}abonly function. This function utilizes numpy multiprocessing
module to split the Gadget II calculations into multiple threads and can also write the
Gadget II values for any adinkra to a text file. It is noticeably different from the original
Gadget equation. This makes the task of calculating all Gadget II values a lengthy
process because there are 36,864 x 36,864 possible combinations of adinkras which
cannot be quickly reduced resulting in a total number of 1,358,954,496 calculations.
Even if the same two adinkras are paired in the equation again but the order is reversed
this does not guarantee the same Gadget value as opposed to the original Gadget equation.

The original Gadget calculation algorithm took $(\mathcal{R})$ adinkra from the set 36,864 adinkras
and then proceeded to calculate the Gadget value for each $\mathcal{G}_{(2)}(\mathcal{
R}, \mathcal{R}')$ pair for n adinkra in set 36,864.  A Python multiprocessing module was
used in calculating the Gadget II values by breaking up the Gadget II calculation from a
single thread into 64 threads. This entailed breaking up the list of 36,864 adinkras into
64 sets so now the $(\mathcal{R})$ adinkra was paired simultatenously across 64 different
threads against 4608 $(\mathcal{R}^\prime)$ adinkras in each thread. Gadget II values were
tracked and counted in each thread and the final Gadget II values for each $(\mathcal{R})$
were written to a dedicated flat file.

The results of these Gadget II apps over the entire library of 36,864 adinkras are shown
in Table 2.

\subsubsection{The ${\rm BC}{}_4$ Small Library Gadget II}
Calculations of Gadget II values for the ${\rm BC}{}_4$ Small Library are done by using the
aforementioned scripts with hardcoded ${\rm BC}{}_4$ Small Library. The Small Library is
defined in fx{\_}bc4small{\_}libdef.py. This script calculates all the 96 adinkras and those
are then fed into fx{\_}vij{\_}holoraumy.py and then fx{\_}gadgets.py. The fx{\_}gadgets.py is
the single threaded version of the fx{\_}mpgadgets.py script. Because there is only 9,216
total Gadget II values for Small Library multithreading is not needed.

The results of these Gadget II apps over the small BC${}_4$ library adinkras are shown
in Table 4.

\subsection{Python Code-II Version}

\subsubsection{Building Libraries}

We begin by constructing all 384 permutation matrices and 16 binary matrices of size \(4\times4\), the dimensions required to describe adinkras in the 1D, $N$ = 4 space. Then, we construct tetrads of the 
signed permutation matrices such that each set of four $\bm {\rm L}$-matrices conforms to the rules required of valid adinkras---these tetrads can be treated similar to adjacency matrices in the construction of each adinkra.

After indexing all 36,864 valid adinkras, we build a dictionary that maps each adinkra to its respective permutation and binary matrices, as well as their resultant $\bm {\rm L}$- and $\bm {\rm R}$-matrices. We then iterate over ${\bm {\rm L}}^{}_\rI$ and ${\bm {\rm R}}^{}_\rI$ for \(I, J = 0, 1, 2, 3\) to make a dictionary mapping our adinkras to their Bosonic Holoraumy matrices (i.e. \({\bm V}\)-matrices). Note,  we note $(\cal R)$ = 1, $\dots$ 36,864 and
\begin{equation}
  \bm{{V}}_{\rI\rJ}{}^{(\cal R)} = -i\frac{1}{2} \left[
{\bm {\rm L}}^{(\cal R)}_\rI \,  {\bm {\rm R}}^{(\cal R)}_\rJ ~-~ 
{\bm {\rm L}}^{(\cal R)}_\rJ \,  {\bm {\rm R}}^{(\cal R)}_\rI  \right]  ~~~.
\end{equation}
We can analogously construct a Fermionic Holoraumy matrix (i.e. $( \bm {\Tilde{V}}$-matrix) dictionary, since
\begin{equation}
\bm{\Tilde {V}}_{\rI\rJ}{}^{(\cal R)} = -i\frac{1}{2} 
 \left[
{\bm {\rm R}}^{(\cal R)}_\rI \,  {\bm {\rm L}}^{(\cal R)}_\rJ ~-~ 
{\bm {\rm R}}^{(\cal R)}_\rJ \,  {\bm {\rm L}}^{(\cal R)}_\rI  \right]
\end{equation}

Saving our dictionaries for quick recollection in future runs of our calculation, we modify our code to read our adinkra and matrix libraries from local files, in lieu of recreating our results every run. This sets the foundation for our calculation of the Second Gadget over the whole space of 1D, $N$ = 4 adinkras.

\subsubsection{Calculating in Python}

The calculation of the Second Gadget is executed after recalling our matrix and adinkra libraries, and is carried out over a user-specified range. 

We define a function that takes two adinkras as input, ``adink1'' and ``adink2.'' The former is given by a for-loop over the specified range of adinkra indices, while the latter is iterated over the the whole space of 36,864 adinkra. 

This function, which we denote as ``gadgetVV,'' explicitly carries out one of the calculations described by 
\ref{Gdgt2zz} by iterating over the indices I and J.

Programmatically, this is done in the form of four nested for-loops. While this approach is not ideal for efficiency, it was taken to simplify the original code and aid in collaboration. Calling gadgetVV over all pairs of adink1 and adink2 (where adink1 lay within the user-specified range) effectively calculates all gadget values that involve the adinkra denoted adink1 in the entire 1.3 billion space of gadget values.

Evaluating gadgetVV over all possible (adink1, adink2) pairs takes an unreasonable amount of time, so we took advantage of Python's support for parallelization. We opted to utilize ``multiprocessing,'' which is a Python package that ``supports spawning processes using an API similar to the threading module'' \cite{PS}. Again we chose to do this instead of utilizing GPU parallelization or rewriting our code in a parallel-friendlier environment due to collaborative considerations.

From the multiprocessing module, we primarily utilized the Pool object, which offers a ``convenient means of parallelizing the execution of a function across multiple input values, distributing the input data across processes (data parallelism)''\cite{PS}.

This allowed us to submit multiple instances of the gadgetVV function that ran in parallel over all of our desired (adink1, adink2) tuples. By splitting up the whole adinkra space over a farm of 64-core CPUs, and parallelizing Gadget calculations over all 64 cores on each machine, we were able to drastically reduce the computation time. 

We also noted that gadgetVV(adink1, adink2) = gadgetVV(adink2, adink1), and that gadgetVV(adink1, adink1) = 0. Thus by limiting the value of adink2 to the range of indices between adink1 and 36,863, we more than halved our overall set of calculations.

\subsection{MATLAB Version}

A MATLAB code was written in order to calculate the values of the second Gadget matrix
over the �small� BC${}_4$ library. This code first imports a data file containing the $\ell $
and $\Tilde \ell$ coefficients of the small library adinkras, in arrays formatted to match their
indices. The superscript indices $({\cal R})$, ranging from 1 to 96, are identified by the row
number of each element in the arrays. The subscript indices IJ, consisting of 12, 13, 14, 23,
24, and 34, are similarly identified by the column number. For each small library coefficient
set with given $({\cal R})$ and IJ indices, only the single nonzero value corresponding to
the appropriate a-index is included, in order to reduce redundant computations.

With these arrays imported, the code uses a nested FOR loop to run Equation 3.4 for each
combination of two adinkras R and R� from the small library. The results, which are the values
of the second Gadget over the small library, are then output as a 96 $\times $ 96 matrix. Lastly,
the code counts the number of occurrences of each the values 0, 1, -1, 1/3, and -1/3 in the
produced matrix.

\newpage
\section{Conclusion}
\label{conclusions}

 \vskip,2in

This work gives an explicit proof of the existence of a second Gadget as well as a description 
of its 1,358,954,496 matrix elements.  Libraries of data of the explicit values of the matrix 
elements of the second Gadget, as well as those of the first Gadget, have been created and
are available upon request.  The existence of these libraries could be used as a foundation
to use ``Big Data'' approaches to cull knowledge from these of data sets.
 
An alternate analytical approach involves the continued study of these data sets.  Of particular
note is the sparse nature of the ${ {\cal G}}{}_{(1)} \left[  ({\cal R}) \, , \,  ({\cal R}^{\prime}) \right]$
and ${ {\cal G}}{}_{(2)} \left[  ({\cal R}) \, , \,  ({\cal R}^{\prime}) \right]$ libraries with approximately
eighty-two percent of their entries being zero.  Of the remaining 2 $\times$ (1,358,954,496)
$\times$ 18 \%
non-vanishing matrix elements in both libraries together, all take on one of the four values, 
$\pm$ 1/3. or $\pm$ 1. This situation is highly suggestive that an analytical approach to be used 
for continued progress involves understanding what symmetry realizations are likely hidden 
within these libraries.    The extremely ordered nature of this result suggests the presence of 
an undiscovered symmetry in the Gadget systems. This is a lesson from the long history of 
quantum theory...when matrix elements vanish in an orderly fashion, a symmetry is usually 
the cause.  This will be the subject of future effort.

 \vspace{.05in}
 \begin{center}
\parbox{4in}{{\it ``All thought is reflection.'' \\ ${~}$ 
\\ ${~}$ }\,\,-\,\, H.\ M.\ S.\ Coxeter $~~~~~~~~~$}
 \parbox{4in}{
 $~~$}  
 \end{center}

 \noindent
{\bf Acknowledgements}\\[.1in] \indent
L.\ Kang, D.\ Kessler and  V.\ Korotkikh 
would like to acknowledge their participation in the second annual
``Brown University Adinkra Math/Phys Hangout'' (18-22 Dec. 
2017).

Additional acknowledgment is given to Prof.\ Charles Doran, Prof. Kevin
Ida, Prof.\ Yan X.\ Zhang, and Dr. Jordan Kostiuk for conversations which 
stimulated efficient techniques for considerations surrounding Gadgets and 
adinkras.

We wish to acknowledge the work of Sylvester J. Gates, III for the creation of 
the 96 $\times$ 96 pixel representation of the matrix elements of the second 
Gadget over the small BC${}_4$ library.

\newpage
$$~~$$

\end{document}